\documentclass[%
 reprint,
 superscriptaddress,
 amsmath,amssymb,
 aps,
 prl,
 longbibliography
]{revtex4-1}

\usepackage{graphicx}
\usepackage{dcolumn}
\usepackage{bm}
\usepackage{svg}
\usepackage{lineno}

\begin{document}


\title{Time-integrated Neutrino Source Searches with 10 years of IceCube Data}
\affiliation{III. Physikalisches Institut, RWTH Aachen University, D-52056 Aachen, Germany}
\affiliation{Department of Physics, University of Adelaide, Adelaide, 5005, Australia}
\affiliation{Dept. of Physics and Astronomy, University of Alaska Anchorage, 3211 Providence Dr., Anchorage, AK 99508, USA}
\affiliation{Dept. of Physics, University of Texas at Arlington, 502 Yates St., Science Hall Rm 108, Box 19059, Arlington, TX 76019, USA}
\affiliation{CTSPS, Clark-Atlanta University, Atlanta, GA 30314, USA}
\affiliation{School of Physics and Center for Relativistic Astrophysics, Georgia Institute of Technology, Atlanta, GA 30332, USA}
\affiliation{Dept. of Physics, Southern University, Baton Rouge, LA 70813, USA}
\affiliation{Dept. of Physics, University of California, Berkeley, CA 94720, USA}
\affiliation{Lawrence Berkeley National Laboratory, Berkeley, CA 94720, USA}
\affiliation{Institut f{\"u}r Physik, Humboldt-Universit{\"a}t zu Berlin, D-12489 Berlin, Germany}
\affiliation{Fakult{\"a}t f{\"u}r Physik {\&} Astronomie, Ruhr-Universit{\"a}t Bochum, D-44780 Bochum, Germany}
\affiliation{Universit{\'e} Libre de Bruxelles, Science Faculty CP230, B-1050 Brussels, Belgium}
\affiliation{Vrije Universiteit Brussel (VUB), Dienst ELEM, B-1050 Brussels, Belgium}
\affiliation{Dept. of Physics, Massachusetts Institute of Technology, Cambridge, MA 02139, USA}
\affiliation{Dept. of Physics and Institute for Global Prominent Research, Chiba University, Chiba 263-8522, Japan}
\affiliation{Dept. of Physics and Astronomy, University of Canterbury, Private Bag 4800, Christchurch, New Zealand}
\affiliation{Dept. of Physics, University of Maryland, College Park, MD 20742, USA}
\affiliation{Dept. of Astronomy, Ohio State University, Columbus, OH 43210, USA}
\affiliation{Dept. of Physics and Center for Cosmology and Astro-Particle Physics, Ohio State University, Columbus, OH 43210, USA}
\affiliation{Niels Bohr Institute, University of Copenhagen, DK-2100 Copenhagen, Denmark}
\affiliation{Dept. of Physics, TU Dortmund University, D-44221 Dortmund, Germany}
\affiliation{Dept. of Physics and Astronomy, Michigan State University, East Lansing, MI 48824, USA}
\affiliation{Dept. of Physics, University of Alberta, Edmonton, Alberta, Canada T6G 2E1}
\affiliation{Erlangen Centre for Astroparticle Physics, Friedrich-Alexander-Universit{\"a}t Erlangen-N{\"u}rnberg, D-91058 Erlangen, Germany}
\affiliation{Physik-department, Technische Universit{\"a}t M{\"u}nchen, D-85748 Garching, Germany}
\affiliation{D{\'e}partement de physique nucl{\'e}aire et corpusculaire, Universit{\'e} de Gen{\`e}ve, CH-1211 Gen{\`e}ve, Switzerland}
\affiliation{Dept. of Physics and Astronomy, University of Gent, B-9000 Gent, Belgium}
\affiliation{Dept. of Physics and Astronomy, University of California, Irvine, CA 92697, USA}
\affiliation{Karlsruhe Institute of Technology, Institut f{\"u}r Kernphysik, D-76021 Karlsruhe, Germany}
\affiliation{Dept. of Physics and Astronomy, University of Kansas, Lawrence, KS 66045, USA}
\affiliation{SNOLAB, 1039 Regional Road 24, Creighton Mine 9, Lively, ON, Canada P3Y 1N2}
\affiliation{Department of Physics and Astronomy, UCLA, Los Angeles, CA 90095, USA}
\affiliation{Department of Physics, Mercer University, Macon, GA 31207-0001, USA}
\affiliation{Dept. of Astronomy, University of Wisconsin, Madison, WI 53706, USA}
\affiliation{Dept. of Physics and Wisconsin IceCube Particle Astrophysics Center, University of Wisconsin, Madison, WI 53706, USA}
\affiliation{Institute of Physics, University of Mainz, Staudinger Weg 7, D-55099 Mainz, Germany}
\affiliation{Department of Physics, Marquette University, Milwaukee, WI, 53201, USA}
\affiliation{Institut f{\"u}r Kernphysik, Westf{\"a}lische Wilhelms-Universit{\"a}t M{\"u}nster, D-48149 M{\"u}nster, Germany}
\affiliation{Bartol Research Institute and Dept. of Physics and Astronomy, University of Delaware, Newark, DE 19716, USA}
\affiliation{Dept. of Physics, Yale University, New Haven, CT 06520, USA}
\affiliation{Dept. of Physics, University of Oxford, Parks Road, Oxford OX1 3PU, UK}
\affiliation{Dept. of Physics, Drexel University, 3141 Chestnut Street, Philadelphia, PA 19104, USA}
\affiliation{Physics Department, South Dakota School of Mines and Technology, Rapid City, SD 57701, USA}
\affiliation{Dept. of Physics, University of Wisconsin, River Falls, WI 54022, USA}
\affiliation{Dept. of Physics and Astronomy, University of Rochester, Rochester, NY 14627, USA}
\affiliation{Oskar Klein Centre and Dept. of Physics, Stockholm University, SE-10691 Stockholm, Sweden}
\affiliation{Dept. of Physics and Astronomy, Stony Brook University, Stony Brook, NY 11794-3800, USA}
\affiliation{Dept. of Physics, Sungkyunkwan University, Suwon 16419, Korea}
\affiliation{Institute of Basic Science, Sungkyunkwan University, Suwon 16419, Korea}
\affiliation{Dept. of Physics and Astronomy, University of Alabama, Tuscaloosa, AL 35487, USA}
\affiliation{Dept. of Astronomy and Astrophysics, Pennsylvania State University, University Park, PA 16802, USA}
\affiliation{Dept. of Physics, Pennsylvania State University, University Park, PA 16802, USA}
\affiliation{Dept. of Physics and Astronomy, Uppsala University, Box 516, S-75120 Uppsala, Sweden}
\affiliation{Dept. of Physics, University of Wuppertal, D-42119 Wuppertal, Germany}
\affiliation{DESY, D-15738 Zeuthen, Germany}

\author{M. G. Aartsen}
\affiliation{Dept. of Physics and Astronomy, University of Canterbury, Private Bag 4800, Christchurch, New Zealand}
\author{M. Ackermann}
\affiliation{DESY, D-15738 Zeuthen, Germany}
\author{J. Adams}
\affiliation{Dept. of Physics and Astronomy, University of Canterbury, Private Bag 4800, Christchurch, New Zealand}
\author{J. A. Aguilar}
\affiliation{Universit{\'e} Libre de Bruxelles, Science Faculty CP230, B-1050 Brussels, Belgium}
\author{M. Ahlers}
\affiliation{Niels Bohr Institute, University of Copenhagen, DK-2100 Copenhagen, Denmark}
\author{M. Ahrens}
\affiliation{Oskar Klein Centre and Dept. of Physics, Stockholm University, SE-10691 Stockholm, Sweden}
\author{C. Alispach}
\affiliation{D{\'e}partement de physique nucl{\'e}aire et corpusculaire, Universit{\'e} de Gen{\`e}ve, CH-1211 Gen{\`e}ve, Switzerland}
\author{K. Andeen}
\affiliation{Department of Physics, Marquette University, Milwaukee, WI, 53201, USA}
\author{T. Anderson}
\affiliation{Dept. of Physics, Pennsylvania State University, University Park, PA 16802, USA}
\author{I. Ansseau}
\affiliation{Universit{\'e} Libre de Bruxelles, Science Faculty CP230, B-1050 Brussels, Belgium}
% ...
\date{\today}

\date{\today}

\pacs{Valid PACS appear here}

\begin{abstract}
 This paper presents the results from point-like neutrino source searches using ten years of IceCube data collected between Apr.~6, 2008 and Jul.~10, 2018. 
 We evaluate the significance of an astrophysical signal from a point-like source looking for an excess of clustered neutrino events with energies typically above $\sim1$\,TeV among the background of atmospheric muons and neutrinos. We perform a full-sky scan, a search within a selected source catalog, a catalog population study, and three stacked Galactic catalog searches. 
The most significant point in the Northern hemisphere from scanning the sky is coincident with the Seyfert II galaxy NGC 1068, which was included in the source catalog search.
The excess at the coordinates of NGC 1068 is inconsistent with background expectations at the level of $2.9\,\sigma$ after accounting for statistical trials. The combination of this result along with excesses observed at the coordinates of three other sources, including TXS 0506+056, suggests that, collectively, correlations with sources in the Northern catalog are inconsistent with background at 3.3$\,\sigma$ significance.
These results, all based on searches for a cumulative neutrino signal integrated over the ten years of available data, motivate further study of these and similar sources, including time-dependent analyses, multimessenger correlations, and the possibility of stronger evidence with coming upgrades to the detector.

\end{abstract}
\maketitle

Cosmic rays (CRs) have been observed for over a hundred years\,\citep{Hess:1912srp} penetrating the entire surface of the Earth's atmosphere in the form of leptonic and hadronic charged particles with energies up to $\sim10^{20}$\,eV\,\citep{Bird:1994mp}. The origin of these particles is still largely unknown since they are deflected on their journey to the Earth by magnetic fields. 
Very-high-energy (VHE) $\gamma$-rays ($E_\gamma>$100\,GeV) travel without deflection and so provide evidence for astrophysical acceleration sites. However, these photons can be produced by both leptonic and hadronic processes and are attenuated by extragalactic background light, meaning they cannot probe distances larger than $z\sim1$ at energies above $\sim$1\,TeV. 
In comparison, only hadronic processes can produce an astrophysical neutrino flux which would travel unattenuated and undeflected from the source to the Earth.  
Thus, astrophysical neutrino observations are critical to identify CR sources, or to discover distant very-high-energy accelerators.

IceCube has discovered astrophysical neutrinos in multiple diffuse flux searches \cite{Aartsen:2016xlq,Aartsen:2017mau,Aartsen2013,Aartsen:2014gkd}. 
Notably, a potential neutrino source, TXS 0506+056, has been identified through a multi-messenger campaign around a high-energy IceCube event in Sep.~2017~\cite{eaat1378}. IceCube also found evidence for neutrino emission over $\sim$110 days from 2014-15 at the location of TXS 0506+056 when examining over 9 years of archival data~\cite{IceCube:2018cha}. 
Nonetheless, the estimated flux from this source alone is less than 1\% of the total astrophysical neutrino flux~\cite{Aartsen:2016xlq}.
In this paper we search for various point-like neutrino sources using 10 years of IceCube observations.

The IceCube neutrino telescope is a cubic kilometer array of digital optical modules (DOMs) each containing a 10'' PMT~\cite{2010NIMPA.139A} and on-board read-out electronics~\cite{Abbasi:2008aa}. These DOMs are arranged in 86 strings between 1.45 and 2.45\,km below the surface of the ice at the South Pole~\citep{Aartsen:2016nxy}. The DOMs are sensitive to Cherenkov light from energy losses of ultra-relativistic charged particles traversing the ice. This analysis targets astrophysical muon neutrinos and antineutrinos ($\nu_\mu$), which undergo charged-current interactions in the ice to produce a muon traversing the detector. The majority of the background for this analysis originates from CRs interacting with the atmosphere to produce showers of particles including atmospheric muons and neutrinos. The atmospheric muons from the Southern hemisphere are able to penetrate the ice and are detected as track-like events in IceCube at a rate orders of magnitude higher than the corresponding atmospheric neutrinos~\cite{Aartsen:2016nxy}. Almost all of the atmospheric muons from the Northern hemisphere are filtered out by the Earth. However, poorly-reconstructed atmospheric muons from the Southern sky create a significant background in the Northern hemisphere.
Atmospheric neutrinos also produce muons from charged-current $\nu_\mu$ interactions, acting as an irreducible background in both hemispheres. 
Neutral-current interactions or $\nu_e$ and $\nu_\tau$ charged-current interactions produce particle showers with spherical morphology known as cascade events. Tracks at $\sim$\,TeV energies are reconstructed with a typical angular resolution of $\lesssim 1^\circ$, while cascades have an angular resolution of $\sim10^\circ-15^\circ$\citep{Aartsen:2017eiu}. This analysis selects track-like events because of their better angular resolution. Tracks have the additional advantage that they can be used even if the neutrino interaction vertex is located outside of the detector. This greatly increases the detectable event rate.

\begin{table}
\caption{IceCube configuration, livetime, number of events, start and end date and published reference in which the sample selection is described. }
\label{tab:livetimesAbr}
\centering
\begin{tabular}{p{1.2cm}  p{1.2cm}  p{1.3cm} p{1.7cm} p{1.7cm} p{0.7cm}} \hline \\[-5pt]
    \multicolumn{6}{c}{Data Samples} \\[5pt]
    \hline
    Year & Livetime (Days) & Number of Events & Start Day & End Day & Ref.\\[5pt] \hline
    IC40 & 376.4 & 36900 & 2008/04/06 & 2009/05/20 & \citep{Abbasi:2010rd} \\[5pt] \hline
    IC59 & 352.6 & 107011 & 2009/05/20 & 2010/05/31 & \citep{Aartsen:2013uuv} \\[5pt] \hline
    IC79 & 316.0 & 93133 & 2010/06/01 & 2011/05/13 & \citep{Schatto:2014kbj} \\[5pt] \hline
    IC86-2011 & 332.9 & 136244 & 2011/05/13 & 2012/05/15 & \citep{Aartsen:2014cva} \\[5pt] \hline
    IC86-2012-18 & 2198.2 & 760923 & 2012/04/26\footnotemark[1] & 2018/07/10 & This work \\[5pt]
    \hline
\end{tabular}
\footnotetext[1]{start date for test runs of the new processing. The remainder of this run began 2012/05/15}
\end{table}

\begin{figure}
    \centering
    \includegraphics[width=0.5\textwidth]{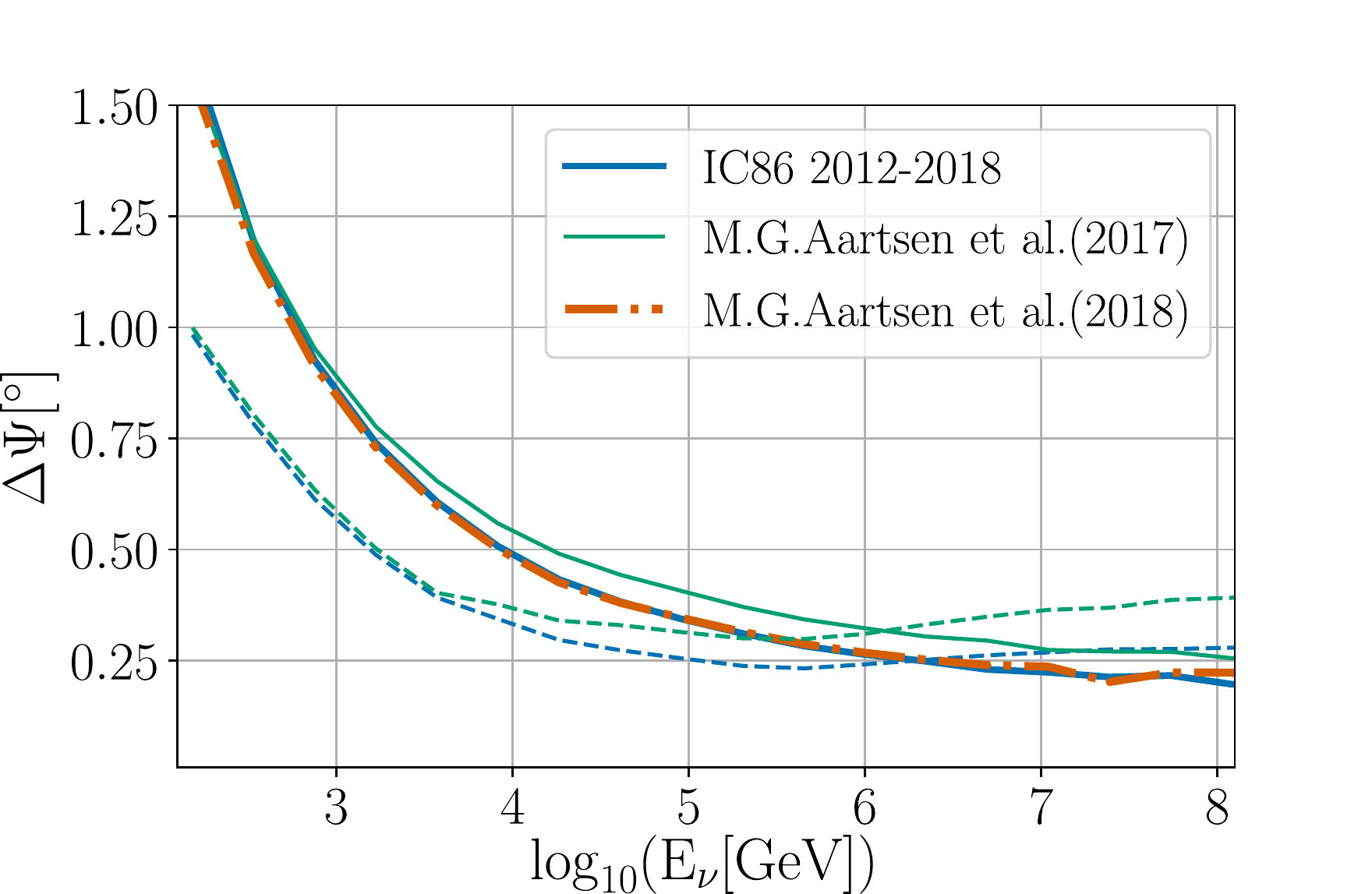}
    \caption{The median angle between simulated neutrino and reconstructed muon directions as a function of energy for the data selection used in the latest 6 years compared to that in Ref.~\cite{Aartsen:2016oji} (solid and dashed lines are for Northern and Southern hemispheres respectively)
    and in Ref.~\cite{Aartsen:2018ywr} for the Northern hemisphere. }
    \label{fig:psf}
\end{figure}

During the first three years of data included here, IceCube was incomplete and functioned with 40, 59, and 79 strings. For these years and also during the first year of data taking of the full detector (IC86), the event selection and reconstruction was updated until it stabilized in 2012, as detailed in Table.~\ref{tab:livetimesAbr}. Seven years of tracks were previously analyzed to search for point sources~\cite{Aartsen:2016oji}. Subsequently, an eight-year sample of tracks from the Northern sky used for diffuse muon neutrino searches was also analyzed looking for point sources~\cite{Aartsen:2018ywr}. The aim of this work is to introduce a selection which unifies the event filtering adopted in these two past searches.
Additionally, the direction reconstruction~\citep{Ahrens:2003fg,Aartsen:2013bfa} has been updated to use the deposited event energy in the detector. This improves the angular resolution by more than 10$\%$ for events above 10\,TeV compared to the seven-year study~\cite{Aartsen:2016oji}, and achieves a similar angular resolution to the eight-year Northern diffuse track selection~\cite{Aartsen:2018ywr} which also uses deposited event energy in the direction reconstruction (see Fig.~\ref{fig:psf}). The absolute pointing accuracy of IceCube has been demonstrated to be $\lesssim0.2^\circ$~\cite{Aartsen:2013zka} via measurements of the effect of the Moon shadow on the background CR flux. 

Different criteria are applied to select track-like events from the Northern and Southern hemisphere (with a boundary between them at declination $\delta=-5^\circ$), because the background differs in these two regions. 
Almost all the atmospheric muons in the Northern hemisphere can be removed by selecting high-quality track-like events. In the Southern hemisphere, the atmospheric background is reduced by strict cuts on the reconstruction quality and minimum energy, since the astrophysical neutrino fluxes are expected to have a harder energy spectrum than the background of atmospheric muons and neutrinos. This effectively removes almost all Southern hemisphere events with an estimated energy below $\sim10$\,TeV (see Fig.~\ref{fig:enPDF} in the supplementary material).

In both hemispheres, atmospheric muons and cascade events are further filtered using multi-variate Boosted Decision Trees (BDT). In this analysis, a single BDT is trained to recognize three classes of events in the Northern hemisphere: single muon tracks from atmospheric and astrophysical neutrinos, atmospheric muons, and cascades, where neutrino-induced tracks are treated as signal. This BDT uses 11 variables related to event topology and reconstruction quality.
The Northern BDT preserves $\sim90\%$ of the atmospheric neutrinos and $\sim0.1\%$ of the atmospheric muons from the initial selection of track-like events, 
also applied in previous muon neutrino searches~\citep{Aartsen:2016oji,Aartsen:2018ywr}. In the Southern hemisphere, the BDT and selection filters are taken from Ref.~\citep{Aartsen:2016oji}. The final all-sky event rate of $\sim2\,$mHz is dominated by muons from atmospheric neutrinos in the Northern hemisphere and by high-energy, well-reconstructed muons in the Southern hemisphere. This updated selection applied to the final 6 years of data shown in Table~\ref{tab:livetimesAbr}. The preceding four years of data are handled exactly as in the past. 

The point-source searches conducted in this paper use the existing maximum-likelihood ratio method which compares the hypothesis of point-like signal plus diffuse background versus a background-only null hypothesis. This technique, described in Refs.~\citep{Abbasi:2010rd,Braun:2008bg}, was also applied in the seven and eight-year point source searches~\cite{Aartsen:2016oji,Aartsen:2018ywr}. The all-sky scan and the selected source catalog searches look for directions which maximize the likelihood-ratio in the Northern and Southern hemisphere separately. Since this analysis assumes point-like sources it has sub-optimal to those with extended neutrino emission regions. 
The sensitivity of this analysis to a neutrino flux with an $E^{-2}$ spectrum, calculated according to \cite{Abbasi:2010rd}, shows a $\sim35\%$ improvement compared to the seven-year all-sky search~\citep{Aartsen:2016oji} due to the longer livetime, updated event selection and reconstruction.  While the sensitivity in the Northern hemisphere is comparable to the eight-year study for an $E^{-2}$ spectrum~\citep{Aartsen:2018ywr}, the analysis presented in this work achieves a $\sim30\%$ improvement in sensitivity to sources with a softer spectrum, such as $E^{-3}$. 

\paragraph{All-Sky Scan:}
The brightest sources of astrophysical neutrinos may differ from the brightest sources observed in the electromagnetic (EM) spectrum. For example, cosmic accelerators can be surrounded by a dense medium which attenuates photons emission while neutrinos could be further generated by cosmic-ray interactions in the medium. For this reason, a general all-sky search for the brightest single point-like neutrino source in each hemisphere is conducted that is unbiased by EM observations. This involves evaluating the signal-over-background likelihood-ratio at a grid of points across the entire sky with a finer spacing ($\sim0.1^\circ \times \sim0.1^\circ$) than the typical event angular uncertainty. The points within 8$^\circ$ of the celestial poles are excluded due to poor statistics and limitations in the background estimation technique. 

At each position on the grid, the likelihood-ratio function is maximized resulting in a maximum test-statistic (TS), a best fit number of astrophysical neutrino events ($\hat{n}_s$), and the spectral index ($\hat{\gamma}$) for an assumed power-law energy spectrum. 
The local pre-trial probability (p-value) of obtaining the given or larger TS value at a certain location from only background is estimated at every grid point by fitting the TS distribution from many background trials with a $\chi^2$ function. Each background trial is obtained from the data themselves by scrambling the right ascension, removing any clustering signal. The location of the most significant p-value in each hemisphere is defined to be the hottest spot. The post-trial probability is estimated by comparing the p-value of the hottest spot in the data with a distribution of hottest spots in the corresponding hemisphere from a large number of background trials.

\begin{figure}
    \centering
    \includegraphics[width=0.4\textwidth]{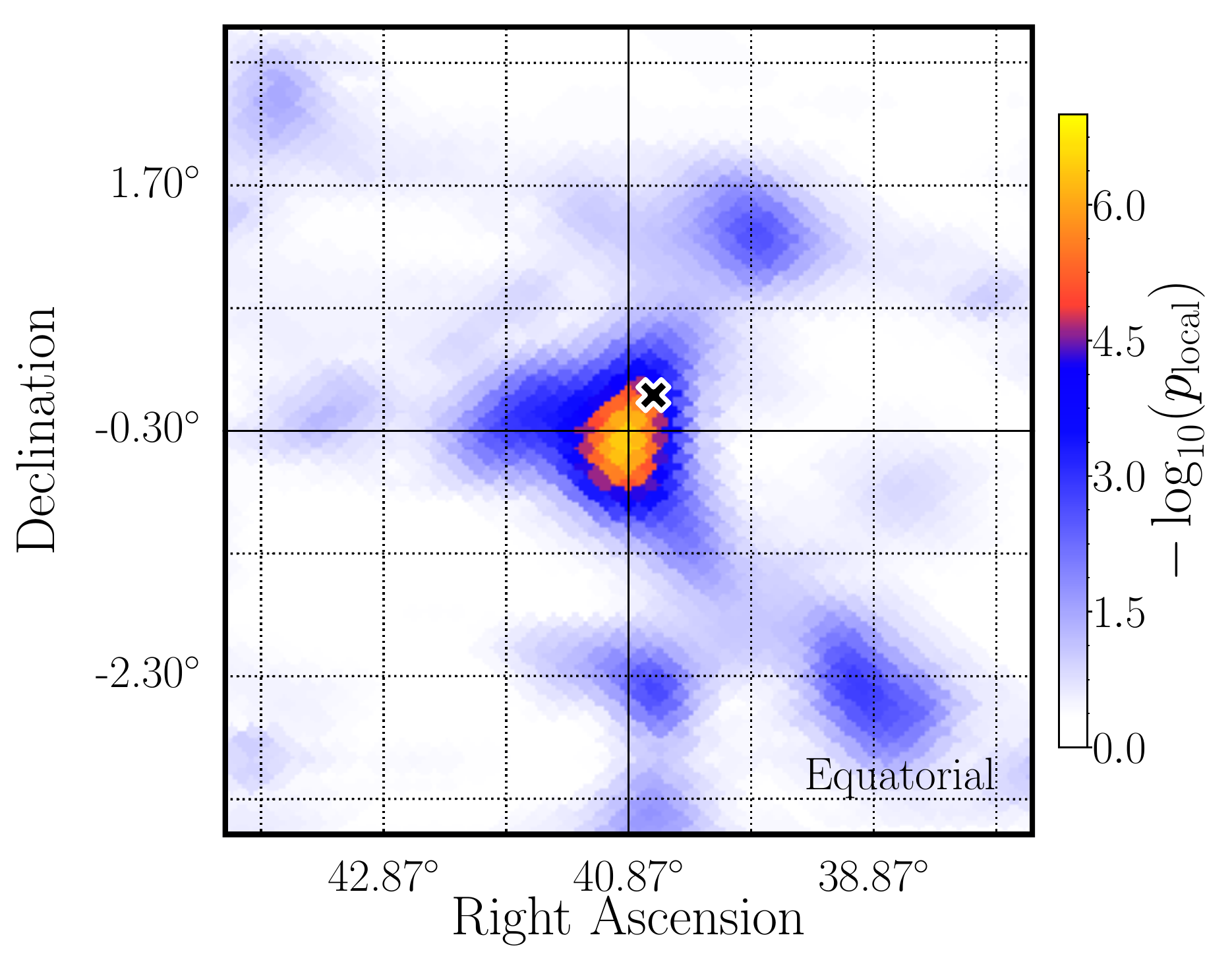}
    \caption{Local pre-trial p-value map around the most significant point in the Northern hemisphere. The black cross marks the coordinates of the galaxy NGC 1068 taken from \textit{Fermi}-4FGL. }
    \label{fig:nHS}
\end{figure}

The most significant point in the Northern hemisphere is found at equatorial coordinates (J2000) right ascension $40.9^\circ$, declination $\text{-}0.3^\circ$ with a local p-value of $3.5\times10^{\text{-}7}$. The best fit parameters at this spot are $\hat{n}_s=61.5$ and $\hat{\gamma}=3.4$. Considering the trials from examining the entire hemisphere reduces this significance to 9.9$\times10^{\text{-}2}$ post-trial.
The probability skymap in a 3$^\circ$ by 3$^\circ$ window around the most significant point in the Northern hemisphere is plotted in Fig.~\ref{fig:nHS}. This point is found 0.35$^\circ$ from the active galaxy NGC 1068, which is also one of the sources in the Northern source catalog. 
The most significant hotspot in the Southern hemisphere, at right ascension $350.2^\circ$ and declination -$56.5^\circ$, is less significant with a pre-trial p-value of $4.3\times10^{\text{-}6}$ and fit parameters $\hat{n}_s=17.8$, and $\hat{\gamma}=3.3$. The significance of this hotspot becomes 0.75 post-trial. Both hotspots alone are consistent with a background-only hypothesis.
\begin{figure}
    \centering
    \includegraphics[width=0.5\textwidth]{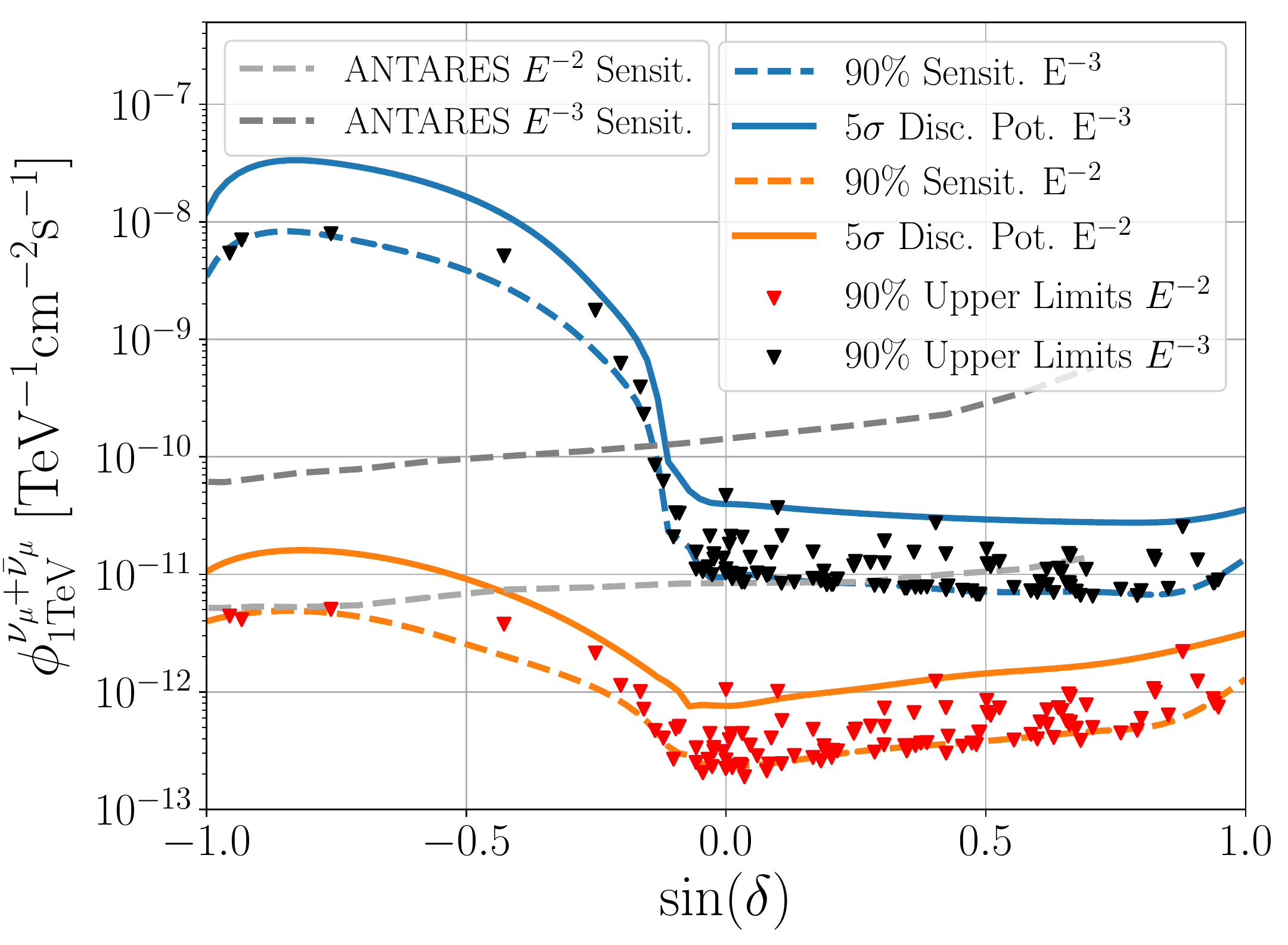}
    \caption{90$\%$ C.L. median sensitivity and 5\,$\sigma$ discovery potential as a function of source declination for a neutrino source with an $E^{-2}$ and $E^{-3}$ spectrum. The 90$\%$ upper-limits are shown excluding an $E^{-2}$ and $E^{-3}$ source spectrum for the sources in the source list.  The grey curves show the 90\% C.L. median sensitivity from 11 yrs of ANTARES data~\cite{Aublin:2019zzn}.}
    \label{fig:UL}
\end{figure}

\paragraph{Source Catalog Searches:} 
The motivation of this search is to improve sensitivity to detect possible neutrino sources already observed in $\gamma$-rays. A new catalog composed of 110 sources has been constructed which updates the catalog used in previous sources searches~\citep{Aartsen:2016oji}. The new catalog uses the latest $\gamma$-ray observations and is based on rigorous application of a few simple criteria, described below. The size of the catalog was chosen to limit the trial factor applied to the most significant source in the catalog such that a 5\,$\sigma$ p-value before trials would remain above 4\,$\sigma$ after trials. These 110 sources are composed of Galactic and extragalactic sources which are selected separately.  

The extragalactic sources are selected from the \textit{Fermi}-LAT 4FGL catalog~\citep{2019arXiv190210045T} since it provides the highest-energy unbiased measurements of $\gamma$-ray sources over the full sky. Sources from 4FGL are weighted according to the integral \textit{Fermi}-LAT flux above 1\,GeV divided by the sensitivity flux for this analysis at the respective source declination. The 5$\%$ highest-weighted BL Lacs and flat spectrum radio quasars (FSRQs) are each selected.  
The minimum weighted integral flux from the combined selection of BL Lac and FSRQs is used as a flux threshold to include sources marked as unidentified blazars and AGN. Eight 4FGL sources are identified as starburst galaxies. Since these types of objects are thought to host hadronic emission~\citep{Loeb:2006tw,Murase:2013rfa}, they are all included in the final source list. 

To select Galactic sources, we consider measurements of VHE $\gamma$-ray sources from TeVCat~\cite{tevcat,2008ICRCTevCat} and gammaCat~\cite{gammacat}. Spectra of the $\gamma$-rays were converted to equivalent neutrino fluxes, assuming a purely hadronic origin of the observed $\gamma$-ray emission where $E_\gamma\simeq2E_\nu$, and compared to the sensitivity of this analysis at the declination of the source (Fig.~\ref{fig:UL}). Those Galactic objects with predicted energy fluxes $>50\%$ of IceCube's sensitivity limit for an $E^{-2}$ spectrum, were included in the source catalog. 
A total of 12 Galactic $\gamma$-ray sources survived the selection.

The final list of neutrino source candidates is a Northern-sky catalog containing 97 objects (87 extragalactic and 10 Galactic) and a Southern-sky catalog containing 13 sources (11 extragalactic and 2 Galactic). The large North-South difference is due to the difference in the sensitivity of IceCube in the Northern and Southern hemispheres.
The post-trial p-value for each catalog describes the significance of the single most significant source in the catalog and is calculated as the fraction of background trials where the pre-trial p-value of the most significant fluctuation is smaller than the pre-trial p-value found in data.

\begin{figure}
    \centering
    \includegraphics[width=0.5\textwidth]{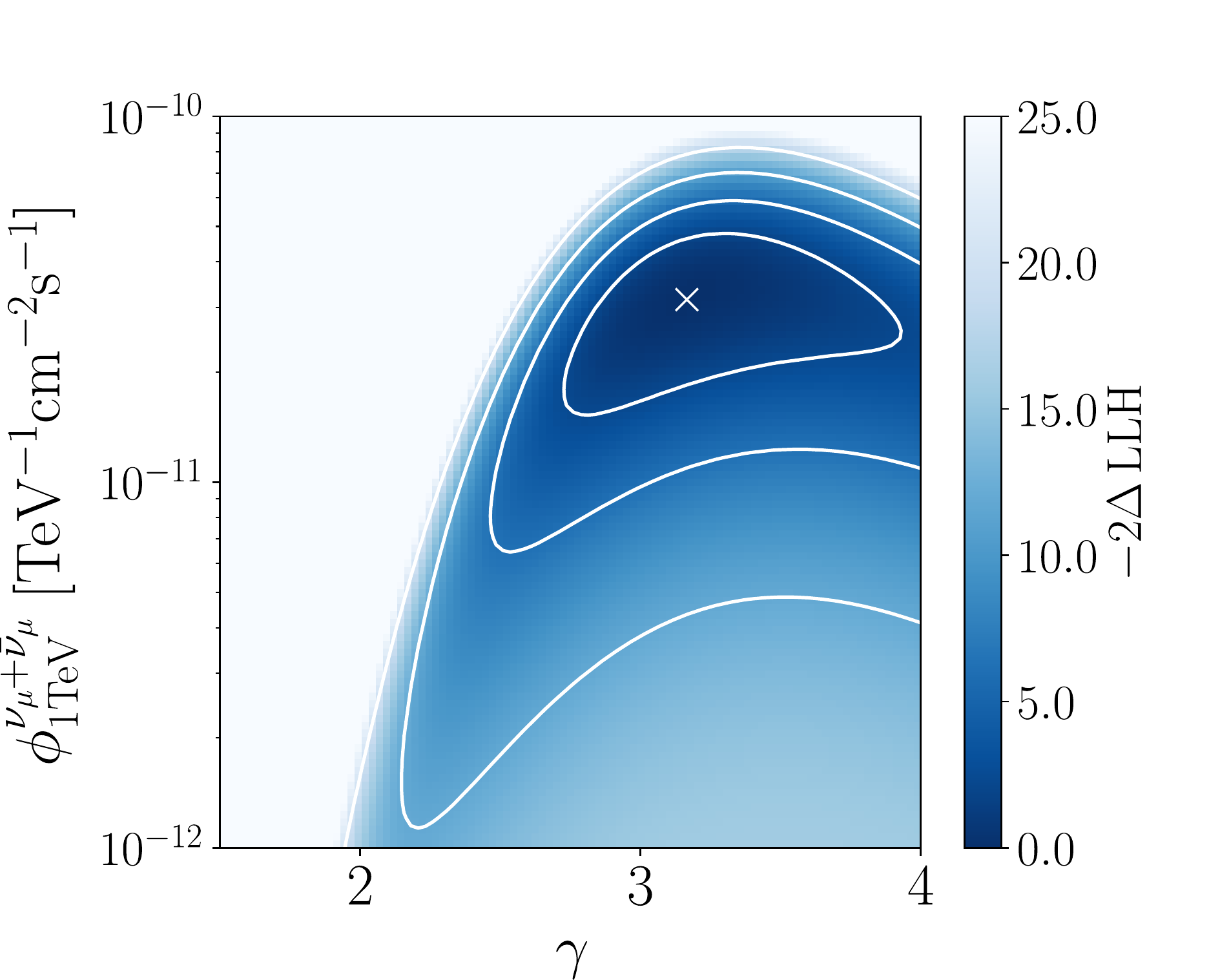}
    \caption{Likelihood map at the position of NGC 1068 as a function of the astrophysical flux spectral index and normalization at 1\,TeV. Contours show 1, 2, 3, and 4\,$\sigma$ confidence intervals assuming Wilks' theorem with 2 degrees of freedom~\citep{Wilks:1938dza}. The best fit spectrum is point marked with ``$\times$". }
    \label{fig:gammaScan}
\end{figure}

The obtained pre-trial p-values are provided in Tab.~\ref{tab:srclist} and their associated 90\% C.L. flux upper-limits are shown in Fig.~\ref{fig:UL}, together with the expected sensitivity and discovery potential fluxes. 
The most significant excess in the Northern catalog of 97 sources is found in the direction of the galaxy NGC 1068, analyzed for the first time by IceCube in this analysis, with a local pre-trial p-value of $1.8\times10^{-5}$ (4.1\,$\sigma$). The best fit parameters are $\gamma=3.2$ and $\hat{n}_s=50.4$, consistent with the results for the all-sky Northern hottest spot, $0.35^\circ$ away. From Fig.~\ref{fig:angErr} and Fig.\ref{fig:nHS} it can be inferred that the significance of the all-sky hotspot and the excess at NGC 1068 are dominated by the same cluster of events. The parameters of the best fit spectrum at the coordinates of NGC 1068 are shown in Fig.~\ref{fig:gammaScan}. 
When the significance of NGC 1068 is compared to the most significant excesses in the Northern catalog from many background trials, the post-trial significance is $2.9\,\sigma$.
To study whether the $0.35^\circ$ offset between the all-sky hotspot and NGC 1068 is typical of the reconstruction uncertainty of a neutrino source, we inject a soft-spectrum source similar to the best-fit $E^{-3.2}$ flux at the position of NGC 1068 in our background samples. Scanning in a $5^\circ$ window around the injection point, we find that the median separation between the most significant hotspot and the injection point is 0.35$^\circ$. 
Thus, if the excess is due to an astrophysical signal from NGC 1068, the offset between the all-sky hotspot and \textit{Fermi}-LAT's coordinates is consistent with the IceCube angular resolution for such a source.

Out of the 13 different source locations examined in the Southern catalog, the most significant excess has a pre-trial p-value of 0.06 in the direction of PKS 2233-148. The associated post-trial p-value is 0.55, which is consistent with background. 

Four sources in the Northern catalog found a pre-trial p-value $<0.01$: NGC 1068, TXS 0506+056, PKS 1424+240, and GB6 J1542+6129. Evidence has been presented for TXS 0506+056 to be a neutrino source~\citep{IceCube:2018cha} using an overlapping event selection in a time-dependent analysis. In this work, in which we only consider the cumulative signal integrated over ten years, we find a pre-trial significance of 3.6\,$\sigma$ at the coordinates of TXS 0506+056 for a best fit spectrum of $E^{-2.1}$, consistent with previous results. 

In addition to the single source search, a source population study is conducted to understand if excesses from several sources, each not yet at evidence level, can cumulatively indicate a population of neutrino sources in the catalog. 

The population study uses the pre-trial p-values of each source in the catalog and searches for an excess in the number of small p-values compared to the uniform background expectation. If the number of objects in the search catalog is $N$, and the number of sources below a given threshold $p_k$ is $k$, then the probability of background producing $k$ or more sources with p-values smaller than $p_k$ is given by the cumulative binomial probability:

\begin{equation}
    p_\mathrm{bkg}=\sum _{i=k}^{N}P_\mathrm{binom}(i|p_k,N)=\sum _{i=k}^{N}\binom{N}{i}p_k^i(1-p_k)^{N-i}\text{ .}
\end{equation}

In order to maximize sensitivity to any possible population size of neutrino sources within the catalog, the probability threshold ($p_k$) is increased iteratively to vary $k$ between 1 and $N$. The result of this search is the most significant $p_{bkg}$ from $N$ different tested values of $k$, then the post-trial p-value from this search must take into account a trial factor for the different tested values of $k$. 

The most significant $p_{bkg}$ from the Northern catalog population analysis is $3.3\times 10^{\text{-}5}$ (4.0$\,\sigma$) which is found when $k=4$ (See Fig.\ref{fig:pop}). The four most significant sources which contribute to this excess are those with p-value $<0.01$ as described above. When accounting for the fact that different signal population sizes are tested, the post-trial p-value is $4.8\times 10^{\text{-}4}$ (3.3$\,\sigma$). 
Since evidence has already been presented for TXS 0506+056 to be a neutrino source~\citep{IceCube:2018cha}, an \textit{a posteriori} search is conducted removing this source from the catalog. The resulting most significant excess is 2.3$\,\sigma$ post-trial due to the remaining three most significant sources. 
For the Southern catalog, the most significant excess is 0.12, provided by 5 of the 13 sources. The resulting post-trial p-value is 0.36.

\paragraph{Stacked Source Searches}
In the case of catalogs of sources that produce similar fluxes, stacking searches require a lower flux per source for a discovery than considering each source individually.  
Three catalogs of Galactic $\gamma$-ray sources are stacked in this paper.
Sources are selected from VHE $\gamma$-ray measurements and categorized into pulsar wind nebulae (PWN), supernova remnants (SNR) and unidentified objects (UNID), with the aim of grouping objects likely to have similar properties as neutrino emitters. The final groups consist of 33 PWN, 23 SNR, and 58 UNID described in Table~\ref{tab:gallist}.
A weighting scheme is adopted to describe the relative contribution expected from each source in a single catalog based on the integral of the extrapolated $\gamma$-ray flux above 10\,TeV. 
All three catalogs find p-values  $>0.1$.

\begin{table}
\caption{Summary of final p-values (pre-trial and post-trial) for each point-like source search implemented in this paper.}
\label{tab:results_summary}
\centering
\begin{tabular}{ p{1.7cm}  p{1.4cm}  p{2.5cm}  p{2.6cm}  }\hline \\[-5pt]
    Analysis    &  Category  & Pre-trial significance ($p_{local}$) &  Post-trial significance \\
    \hline \hline
    All-Sky
    &  North & $3.5\times10^{-7}$ & $9.9\times10^{-2}$ \\
    Scan &  South & $4.3\times10^{-6}$ & 0.75 \\
    \hline
    Source List &  North & $1.8\times10^{-5}$ & $2.0\times10^{-3}$ (2.9$\,\sigma$)\\
    &  South & $5.9\times10^{-2}$ & 0.55 \\
    \hline
    Catalog  & North & 3.3$\times10^{-5}$ & $4.8\times10^{-4}$ (3.3$\,\sigma$) \\
    Population & South & 0.12 & 0.36 \\
    \hline
    Stacking  & SNR & -- & 0.11 \\
    Search& PWN & -- & 1.0 \\
    & UNID & -- & 0.4 \\
    \hline
\end{tabular}
\end{table}

\paragraph{Conclusion}
This paper presents an updated event selection optimized for point-like neutrino source signals applied to 10 years of IceCube data taken from April 2008 to July 2018. Multiple neutrino source searches are performed: an all-sky scan, a source catalog and corresponding catalog population study for each hemisphere, and 3 stacked Galactic-source searches. 

The results of these analyses, all searching for cumulative neutrino signals integrated over the 10 years of data-taking, are summarized in Table~\ref{tab:results_summary}. The most significant source in the Northern catalog, NGC 1068, is inconsistent with a background-only hypothesis at 2.9$\,\sigma$ due to being located 0.35$^\circ$ from the most significant excess in the Northern hemisphere and the Northern source catalog provides a 3.3$\,\sigma$ inconsistency with a background-only hypothesis for the entire catalog. This result comes from an excess of significant p-values in the directions of the Seyfert II galaxy NGC 1068, the blazar TXS 0506+056, and the BL Lacs PKS 1424+240 and GB6 J1542+6129.
NGC 1068, at a 14.4 Mpc distance, is the most luminous Seyfert II galaxy detected by \textit{Fermi}-LAT \cite{2012ApJ...755..164A}. NGC 1068 is an observed particle accelerator, charged particles are accelerated in the jet of the AGN or in the AGN-driven molecular wind~\cite{2016A&A...596A..68L}, producing $\gamma$-rays and potentially neutrinos. Other work has previously indicated NGC 1068 as a potential CR accelerator \cite{2014ApJ...780..137Y,Lacki:2010vs,Loeb:2006tw}. 
Assuming that the observed excess is indeed of astrophysical origin and connected with NGC 1068, the best-fit neutrino spectrum inferred from this work is significantly higher than that predicted from models developed to explain the \emph{Fermi}-LAT gamma-ray measurements (see Fig.~\ref{fig:ngc1068_mwl}). However,
the large uncertainty from our spectral measurement and the high X-ray and $\gamma$-ray absorption along the line of sight~\cite{2015A&A...584A..20W,Lamastra:2017iyo} prevent a straight forward connection.
Time-dependent analyses and the possibility of correlating with multimessenger observations for this and other sources may provide additional evidence of neutrino emission and insights into its origin. Continued data-taking, more refined event reconstruction, and the planned upgrade of IceCube promise further improvements in sensitivity~\cite{vanSanten:2017chb}.
\begin{acknowledgements}
USA {\textendash} U.S. National Science Foundation-Office of Polar Programs,
U.S. National Science Foundation-Physics Division,
Wisconsin Alumni Research Foundation,
Center for High Throughput Computing (CHTC) at the University of Wisconsin-Madison,
Open Science Grid (OSG),
Extreme Science and Engineering Discovery Environment (XSEDE),
U.S. Department of Energy-National Energy Research Scientific Computing Center,
Particle astrophysics research computing center at the University of Maryland,
Institute for Cyber-Enabled Research at Michigan State University,
and Astroparticle physics computational facility at Marquette University;
Belgium {\textendash} Funds for Scientific Research (FRS-FNRS and FWO),
FWO Odysseus and Big Science programmes,
and Belgian Federal Science Policy Office (Belspo);
Germany {\textendash} Bundesministerium f{\"u}r Bildung und Forschung (BMBF),
Deutsche Forschungsgemeinschaft (DFG),
Helmholtz Alliance for Astroparticle Physics (HAP),
Initiative and Networking Fund of the Helmholtz Association,
Deutsches Elektronen Synchrotron (DESY),
and High Performance Computing cluster of the RWTH Aachen;
Sweden {\textendash} Swedish Research Council,
Swedish Polar Research Secretariat,
Swedish National Infrastructure for Computing (SNIC),
and Knut and Alice Wallenberg Foundation;
Australia {\textendash} Australian Research Council;
Canada {\textendash} Natural Sciences and Engineering Research Council of Canada,
Calcul Qu{\'e}bec, Compute Ontario, Canada Foundation for Innovation, WestGrid, and Compute Canada;
Denmark {\textendash} Villum Fonden, Danish National Research Foundation (DNRF), Carlsberg Foundation;
New Zealand {\textendash} Marsden Fund;
Japan {\textendash} Japan Society for Promotion of Science (JSPS)
and Institute for Global Prominent Research (IGPR) of Chiba University;
Korea {\textendash} National Research Foundation of Korea (NRF);
Switzerland {\textendash} Swiss National Science Foundation (SNSF);
United Kingdom {\textendash} Department of Physics, University of Oxford.
\end{acknowledgements}
\bibliography{multimessenger} 

\newpage
\onecolumngrid
\appendix
\section*{Supplementary Material}
\setlength{\tabcolsep}{6pt}

\begin{table}
 \caption{Northern and Southern catalogs used in the \emph{a priori} defined source-list searches. For each source: equatorial coordinates (J2000) from 4FGL are given with the likelihood search results: best-fit number of astrophysical neutrino events $\hat{n}_s$, best-fit astrophysical power-law spectral index $\hat{\gamma}$, local pre-trial p-value -$\log_{10}(p_{local})$, 90\% CL astrophysical flux upper-limit ($\phi_{90\%}$). 
 The neutrino 90\% CL flux upper-limit ($\phi_{90\%}$) is parametrized as: $\frac{dN_{\nu_\mu+\bar{\nu}_\mu}}{dE_\nu}=\phi_{90\%}\cdot\Big(\frac{E_\nu}{TeV}\Big)^{-2}\times10^{-13}\text{TeV}^{-1}\text{cm}^{-2}\text{s}^{-1}$. 
 The four most significant sources with pre-trial p-values less than 0.01 are highlighted in \textbf{bold}. The sources are divided into Northern and Southern catalogs with a boundary at -5$^\circ$ in declination.
 \label{tab:srclist}}
\begin{tabular}[c]{| c  c  c  c  c  c  c  c|}
 \hline
 \multicolumn{8}{ c }{Source List Results}\\
 \hline
 Name & Class & $\alpha\,[\mathrm{deg}]$ & $\delta\,[\mathrm{deg}]$ & $\hat{n}_s$ & $\hat{\gamma}$ & -$\log_{10}(p_{local})$ & $\phi_{90\%}$\\
 \hline

 

 
 
 

PKS 2320-035 & FSRQ & 350.88 & -3.29 & 4.8 & 3.6 & 0.45 & 3.3 \\
3C 454.3 & FSRQ & 343.50 & 16.15 & 5.4 & 2.2 & 0.62 & 5.1 \\
TXS 2241+406 & FSRQ & 341.06 & 40.96 & 3.8 & 3.8 & 0.42 & 5.6 \\
RGB J2243+203 & BLL & 340.99 & 20.36 & 0.0 & 3.0 & 0.33 & 3.1 \\
CTA 102 & FSRQ & 338.15 & 11.73 & 0.0 & 2.7 & 0.30 & 2.8 \\
BL Lac & BLL & 330.69 & 42.28 & 0.0 & 2.7 & 0.31 & 4.9 \\
OX 169 & FSRQ & 325.89 & 17.73 & 2.0 & 1.7 & 0.69 & 5.1 \\
B2 2114+33 & BLL & 319.06 & 33.66 & 0.0 & 3.0 & 0.30 & 3.9 \\
PKS 2032+107 & FSRQ & 308.85 & 10.94 & 0.0 & 2.4 & 0.33 & 3.2 \\
2HWC J2031+415 & GAL & 307.93 & 41.51 & 13.4 & 3.8 & 0.97 & 9.2 \\
Gamma Cygni & GAL & 305.56 & 40.26 & 7.4 & 3.7 & 0.59 & 6.9 \\
MGRO J2019+37 & GAL & 304.85 & 36.80 & 0.0 & 3.1 & 0.33 & 4.0 \\
MG2 J201534+3710 & FSRQ & 303.92 & 37.19 & 4.4 & 4.0 & 0.40 & 5.6 \\
MG4 J200112+4352 & BLL & 300.30 & 43.89 & 6.1 & 2.3 & 0.67 & 7.8 \\
1ES 1959+650 & BLL & 300.01 & 65.15 & 12.6 & 3.3 & 0.77 & 12.3 \\
1RXS J194246.3+1 & BLL & 295.70 & 10.56 & 0.0 & 2.7 & 0.33 & 2.6 \\
RX J1931.1+0937 & BLL & 292.78 & 9.63 & 0.0 & 2.9 & 0.29 & 2.8 \\
NVSS J190836-012 & UNIDB & 287.20 & -1.53 & 0.0 & 2.9 & 0.22 & 2.3 \\
MGRO J1908+06 & GAL & 287.17 & 6.18 & 4.2 & 2.0 & 1.42 & 5.7 \\
TXS 1902+556 & BLL & 285.80 & 55.68 & 11.7 & 4.0 & 0.85 & 9.9 \\
HESS J1857+026 & GAL & 284.30 & 2.67 & 7.4 & 3.1 & 0.53 & 3.5 \\
GRS 1285.0 & UNIDB & 283.15 & 0.69 & 1.7 & 3.8 & 0.27 & 2.3 \\
HESS J1852-000 & GAL & 283.00 & 0.00 & 3.3 & 3.7 & 0.38 & 2.6 \\
HESS J1849-000 & GAL & 282.26 & -0.02 & 0.0 & 3.0 & 0.28 & 2.2 \\
HESS J1843-033 & GAL & 280.75 & -3.30 & 0.0 & 2.8 & 0.31 & 2.5 \\
OT 081 & BLL & 267.87 & 9.65 & 12.2 & 3.2 & 0.73 & 4.8 \\
S4 1749+70 & BLL & 267.15 & 70.10 & 0.0 & 2.5 & 0.37 & 8.0 \\
1H 1720+117 & BLL & 261.27 & 11.88 & 0.0 & 2.7 & 0.30 & 3.2 \\
PKS 1717+177 & BLL & 259.81 & 17.75 & 19.8 & 3.6 & 1.32 & 7.3 \\
Mkn 501 & BLL & 253.47 & 39.76 & 10.3 & 4.0 & 0.61 & 7.3 \\
4C +38.41 & FSRQ & 248.82 & 38.14 & 4.2 & 2.3 & 0.66 & 7.0 \\
PG 1553+113 & BLL & 238.93 & 11.19 & 0.0 & 2.8 & 0.32 & 3.2 \\
\textbf{GB6 J1542+6129} & \textbf{BLL} & \textbf{235.75} & \textbf{61.50} & \textbf{29.7} & \textbf{3.0} & \textbf{2.74} & \textbf{22.0} \\
B2 1520+31 & FSRQ & 230.55 & 31.74 & 7.1 & 2.4 & 0.83 & 7.3 \\
PKS 1502+036 & AGN & 226.26 & 3.44 & 0.0 & 2.7 & 0.28 & 2.9 \\
PKS 1502+106 & FSRQ & 226.10 & 10.50 & 0.0 & 3.0 & 0.33 & 2.6 \\
PKS 1441+25 & FSRQ & 220.99 & 25.03 & 7.5 & 2.4 & 0.94 & 7.3 \\
\textbf{PKS 1424+240} & \textbf{BLL} & \textbf{216.76} & \textbf{23.80} & \textbf{41.5} & \textbf{3.9} & \textbf{2.80} & \textbf{12.3} \\
NVSS J141826-023 & BLL & 214.61 & -2.56 & 0.0 & 3.0 & 0.25 & 2.0 \\
B3 1343+451 & FSRQ & 206.40 & 44.88 & 0.0 & 2.8 & 0.32 & 5.0 \\
S4 1250+53 & BLL & 193.31 & 53.02 & 2.2 & 2.5 & 0.39 & 5.9 \\
PG 1246+586 & BLL & 192.08 & 58.34 & 0.0 & 2.8 & 0.35 & 6.4 \\
MG1 J123931+0443 & FSRQ & 189.89 & 4.73 & 0.0 & 2.6 & 0.28 & 2.4 \\
M 87 & AGN & 187.71 & 12.39 & 0.0 & 2.8 & 0.29 & 3.1 \\
ON 246 & BLL & 187.56 & 25.30 & 0.9 & 1.7 & 0.37 & 4.2 \\
3C 273 & FSRQ & 187.27 & 2.04 & 0.0 & 3.0 & 0.28 & 1.9 \\
4C +21.35 & FSRQ & 186.23 & 21.38 & 0.0 & 2.6 & 0.32 & 3.5 \\
W Comae & BLL & 185.38 & 28.24 & 0.0 & 3.0 & 0.32 & 3.7 \\
PG 1218+304 & BLL & 185.34 & 30.17 & 11.1 & 3.9 & 0.70 & 6.7 \\
PKS 1216-010 & BLL & 184.64 & -1.33 & 6.9 & 4.0 & 0.45 & 3.1 \\
B2 1215+30 & BLL & 184.48 & 30.12 & 18.6 & 3.4 & 1.09 & 8.5 \\
Ton 599 & FSRQ & 179.88 & 29.24 & 0.0 & 2.2 & 0.29 & 4.5 \\
\hline
\end{tabular}
\end{table}

\begin{table}
\begin{tabular}[c]{| c  c  c  c  c  c  c  c|}
 \hline
 Name & Class & $\alpha\,[\mathrm{deg}]$ & $\delta\,[\mathrm{deg}]$ & $\hat{n}_s$ & $\hat{\gamma}$ & -$\log_{10}(p_{local})$ & $\phi_{90\%}$\\
 \hline
PKS B1130+008 & BLL & 173.20 & 0.58 & 15.8 & 4.0 & 0.96 & 4.4 \\
Mkn 421 & BLL & 166.12 & 38.21 & 2.1 & 1.9 & 0.38 & 5.3 \\
4C +01.28 & BLL & 164.61 & 1.56 & 0.0 & 2.9 & 0.26 & 2.4 \\
1H 1013+498 & BLL & 153.77 & 49.43 & 0.0 & 2.6 & 0.29 & 4.5 \\
4C +55.17 & FSRQ & 149.42 & 55.38 & 11.9 & 3.3 & 1.02 & 10.6 \\
M 82 & SBG & 148.95 & 69.67 & 0.0 & 2.6 & 0.36 & 8.8 \\
PMN J0948+0022 & AGN & 147.24 & 0.37 & 9.3 & 4.0 & 0.76 & 3.9 \\
OJ 287 & BLL & 133.71 & 20.12 & 0.0 & 2.6 & 0.32 & 3.5 \\
PKS 0829+046 & BLL & 127.97 & 4.49 & 0.0 & 2.9 & 0.28 & 2.1 \\
S4 0814+42 & BLL & 124.56 & 42.38 & 0.0 & 2.3 & 0.30 & 4.9 \\
OJ 014 & BLL & 122.87 & 1.78 & 16.1 & 4.0 & 0.99 & 4.4 \\
1ES 0806+524 & BLL & 122.46 & 52.31 & 0.0 & 2.8 & 0.31 & 4.7 \\
PKS 0736+01 & FSRQ & 114.82 & 1.62 & 0.0 & 2.8 & 0.26 & 2.4 \\
PKS 0735+17 & BLL & 114.54 & 17.71 & 0.0 & 2.8 & 0.30 & 3.5 \\
4C +14.23 & FSRQ & 111.33 & 14.42 & 8.5 & 2.9 & 0.60 & 4.8 \\
S5 0716+71 & BLL & 110.49 & 71.34 & 0.0 & 2.5 & 0.38 & 7.4 \\
PSR B0656+14 & GAL & 104.95 & 14.24 & 8.4 & 4.0 & 0.51 & 4.4 \\
1ES 0647+250 & BLL & 102.70 & 25.06 & 0.0 & 2.9 & 0.27 & 3.0 \\
B3 0609+413 & BLL & 93.22 & 41.37 & 1.8 & 1.7 & 0.42 & 5.3 \\
Crab nebula & GAL & 83.63 & 22.01 & 1.1 & 2.2 & 0.31 & 3.7 \\
OG +050 & FSRQ & 83.18 & 7.55 & 0.0 & 3.2 & 0.28 & 2.9 \\
TXS 0518+211 & BLL & 80.44 & 21.21 & 15.7 & 3.8 & 0.92 & 6.6 \\
\textbf{TXS 0506+056} & \textbf{BLL} & \textbf{77.35} & \textbf{5.70} & \textbf{12.3} & \textbf{2.1} & \textbf{3.72} & \textbf{10.1} \\
PKS 0502+049 & FSRQ & 76.34 & 5.00 & 11.2 & 3.0 & 0.66 & 4.1 \\
S3 0458-02 & FSRQ & 75.30 & -1.97 & 5.5 & 4.0 & 0.33 & 2.7 \\
PKS 0440-00 & FSRQ & 70.66 & -0.29 & 7.6 & 3.9 & 0.46 & 3.1 \\
MG2 J043337+2905 & BLL & 68.41 & 29.10 & 0.0 & 2.7 & 0.28 & 4.5 \\
PKS 0422+00 & BLL & 66.19 & 0.60 & 0.0 & 2.9 & 0.27 & 2.3 \\
PKS 0420-01 & FSRQ & 65.83 & -1.33 & 9.3 & 4.0 & 0.52 & 3.4 \\
PKS 0336-01 & FSRQ & 54.88 & -1.77 & 15.5 & 4.0 & 0.99 & 4.4 \\
NGC 1275 & AGN & 49.96 & 41.51 & 3.6 & 3.1 & 0.41 & 5.5 \\
\textbf{NGC 1068} & \textbf{SBG} & \textbf{40.67} & \textbf{-0.01} & \textbf{50.4} & \textbf{3.2} & \textbf{4.74} & \textbf{10.5} \\
PKS 0235+164 & BLL & 39.67 & 16.62 & 0.0 & 3.0 & 0.28 & 3.1 \\
4C +28.07 & FSRQ & 39.48 & 28.80 & 0.0 & 2.8 & 0.30 & 3.6 \\
3C 66A & BLL & 35.67 & 43.04 & 0.0 & 2.8 & 0.30 & 3.9 \\
B2 0218+357 & FSRQ & 35.28 & 35.94 & 0.0 & 3.1 & 0.33 & 4.3 \\
PKS 0215+015 & FSRQ & 34.46 & 1.74 & 0.0 & 3.2 & 0.27 & 2.3 \\
MG1 J021114+1051 & BLL & 32.81 & 10.86 & 1.6 & 1.7 & 0.43 & 3.5 \\
TXS 0141+268 & BLL & 26.15 & 27.09 & 0.0 & 2.5 & 0.31 & 3.5 \\
B3 0133+388 & BLL & 24.14 & 39.10 & 0.0 & 2.6 & 0.28 & 4.1 \\
NGC 598 & SBG & 23.52 & 30.62 & 11.4 & 4.0 & 0.63 & 6.3 \\
S2 0109+22 & BLL & 18.03 & 22.75 & 2.0 & 3.1 & 0.30 & 3.7 \\
4C +01.02 & FSRQ & 17.16 & 1.59 & 0.0 & 3.0 & 0.26 & 2.4 \\
M 31 & SBG & 10.82 & 41.24 & 11.0 & 4.0 & 1.09 & 9.6 \\
PKS 0019+058 & BLL & 5.64 & 6.14 & 0.0 & 2.9 & 0.29 & 2.4 \\
\hline
\hline
PKS 2233-148 & BLL & 339.14 & -14.56 & 5.3 & 2.8 & 1.26 & 21.4 \\
HESS J1841-055 & GAL & 280.23 & -5.55 & 3.6 & 4.0 & 0.55 & 4.8 \\
HESS J1837-069 & GAL & 279.43 & -6.93 & 0.0 & 2.8 & 0.30 & 4.0 \\
PKS 1510-089 & FSRQ & 228.21 & -9.10 & 0.1 & 1.7 & 0.41 & 7.1 \\
PKS 1329-049 & FSRQ & 203.02 & -5.16 & 6.1 & 2.7 & 0.77 & 5.1 \\
NGC 4945 & SBG & 196.36 & -49.47 & 0.3 & 2.6 & 0.31 & 50.2 \\
3C 279 & FSRQ & 194.04 & -5.79 & 0.3 & 2.4 & 0.20 & 2.7 \\
PKS 0805-07 & FSRQ & 122.07 & -7.86 & 0.0 & 2.7 & 0.31 & 4.7 \\
PKS 0727-11 & FSRQ & 112.58 & -11.69 & 1.9 & 3.5 & 0.59 & 11.4 \\
LMC & SBG & 80.00 & -68.75 & 0.0 & 3.1 & 0.36 & 41.1 \\
SMC & SBG & 14.50 & -72.75 & 0.0 & 2.4 & 0.37 & 44.1 \\
PKS 0048-09 & BLL & 12.68 & -9.49 & 3.9 & 3.3 & 0.87 & 10.0 \\
NGC 253 & SBG & 11.90 & -25.29 & 3.0 & 4.0 & 0.75 & 37.7 \\
\hline
\hline
\end{tabular}
\end{table}

\begin{figure}
    \centering
    \includegraphics[width=0.44\textwidth]{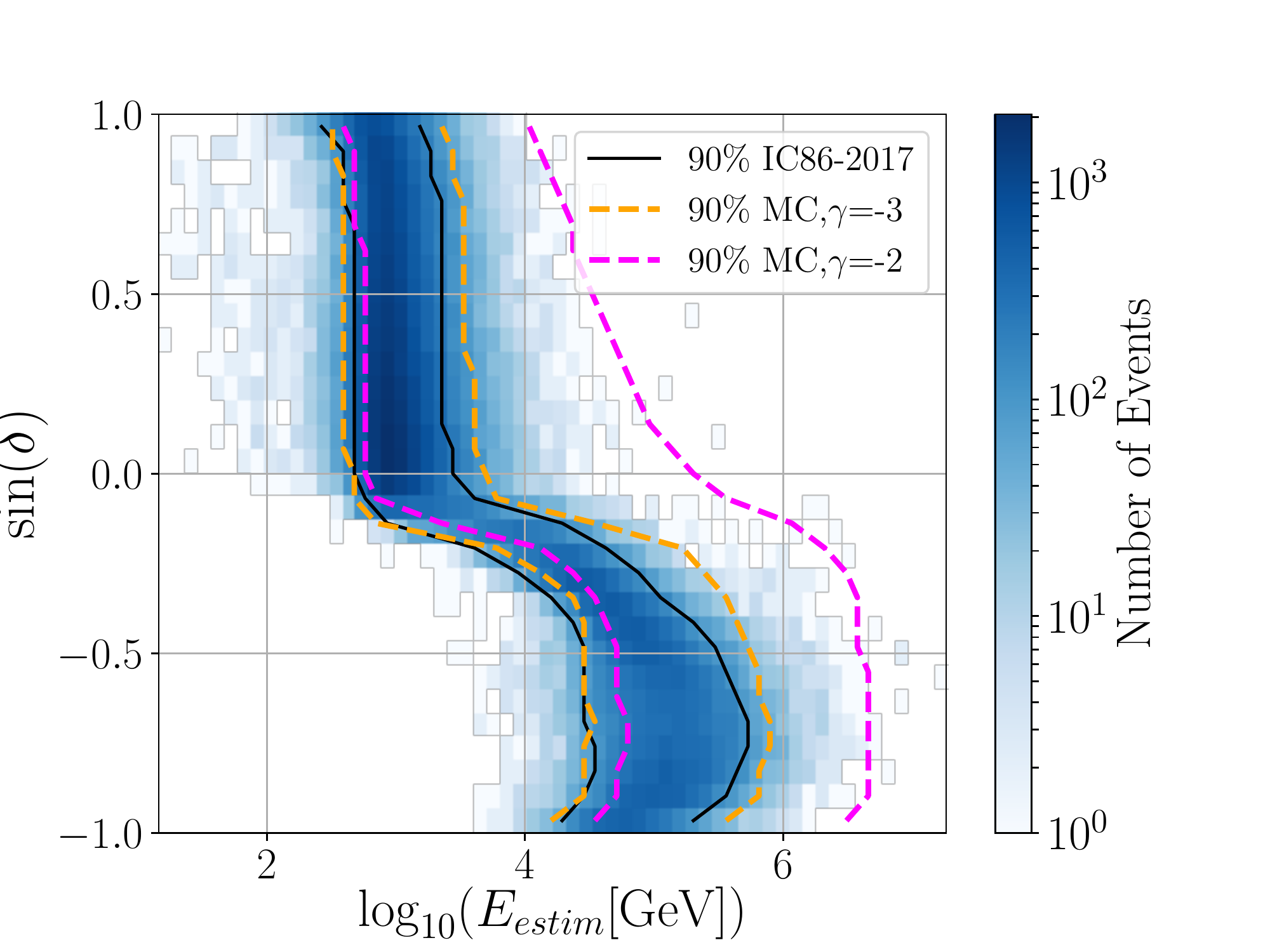}
    \includegraphics[width=0.48\textwidth]{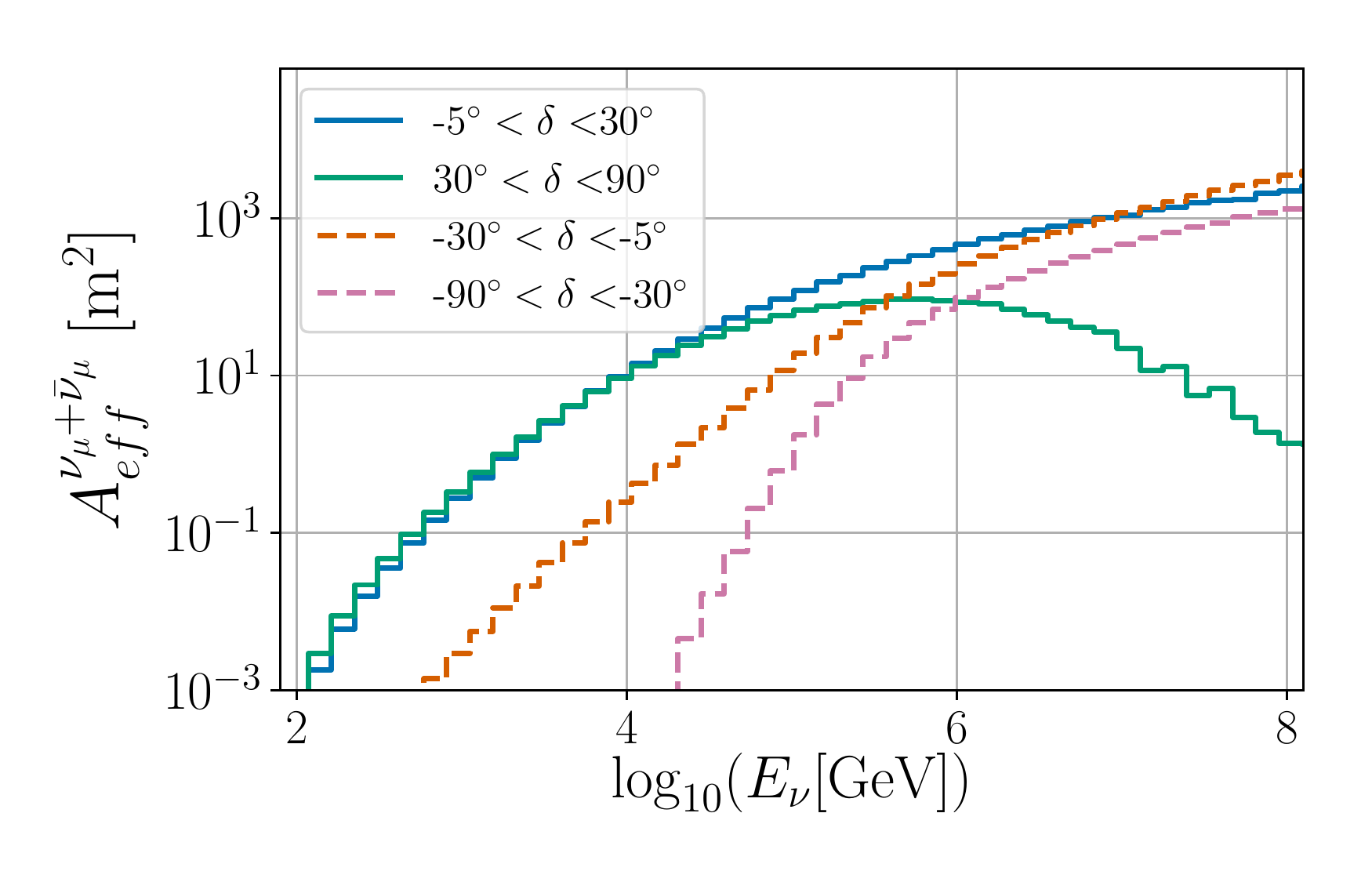}
    \caption{\textit{Left:} The 2D distribution of events in one year of data for the final event selection as a function of reconstructed declination and estimated energy. The 90\% energy range for the data (black), as well as simulated astrophysical signal Monte-Carlo (MC) for an $E^{-2}$ and an $E^{-3}$ spectrum are shown in magenta and orange respectively as a guide for the relevant energy range of IceCube. \textit{Right:} The effective area as a function of neutrino energy for the IC86 2012-2018 event selection averaged across the declination band for several declination bins using simulated data.}
    \label{fig:enPDF}
\end{figure}

The effective area for this search corresponds to the efficiency of the analysis cuts and detector effects to observe an astrophysical neutrino flux as a function of energy and declination. The expected rate of muon neutrinos and anti-neutrinos ($\frac{dN_{\nu+\bar{\nu}}}{dt}$) from a point-like source at declination $\delta$ from a flux ($\phi_{\nu+\bar{\nu}}$) as a function of neutrino energy ($E_\nu$) is:
\begin{equation}
    \frac{dN_{\nu+\bar{\nu}}}{dt}=\int_0^\infty A_{eff}^{\nu+\bar{\nu}}(E_\nu, \delta)\times \phi_{\nu+\bar{\nu}}(E_\nu)\text dE_\nu { .}
\end{equation}
The resulting effective area for the IC86 2012-2018 event selection is shown in Fig.~\ref{fig:enPDF} as a function of simulated neutrino energy in declination bins. The combination of the effective area, angular resolution shown in Fig~\ref{fig:psf}, and the background data rate, determines the analysis sensitivity to a point-like neutrino source. 

\begin{figure}[ht]
    \includegraphics[width=0.65\textwidth]{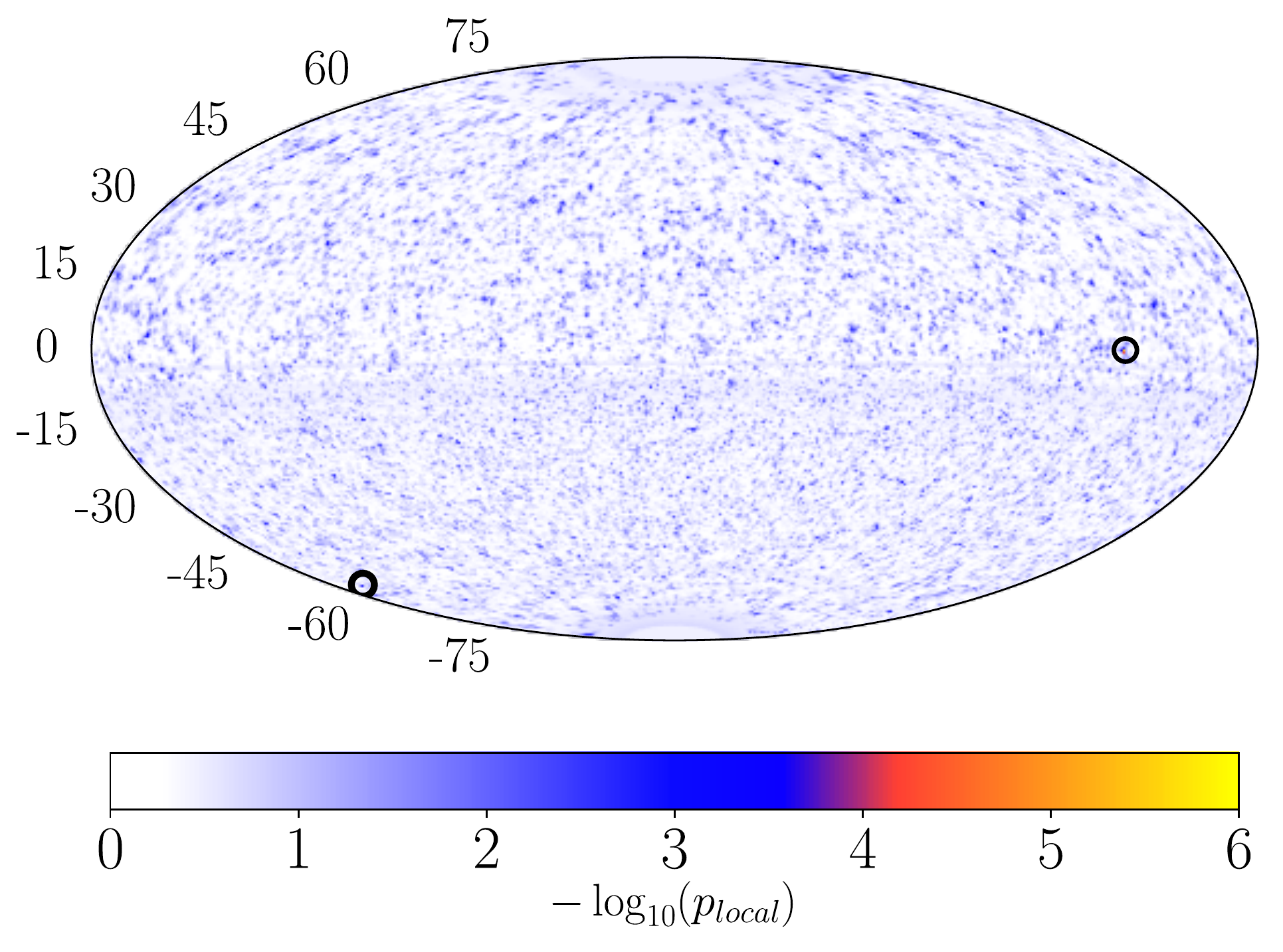}
     \caption{Skymap of -$\log_{10}(p_{local})$, where $p_{local}$ is the local pre-trial p-value, for the sky between $\pm82^\circ$ declination in equatorial coordinates. The Northern and Southern hemisphere hotspots, defined as the most significant $p_{local}$ in that hemisphere, are indicated with black circles. 
     }
    \label{fig:skymap}
\end{figure}

The updated event selection is used to scan each hemisphere for the single most significant point-like neutrino source, and in addition to examine individual sources observed in $\gamma$-rays via the analyses described above. The result of the all-sky scan is discussed above and can be seen in Fig.\ref{fig:skymap}. The details of the source list and the individual results from examining each of the sources in the Northern and Southern catalogs (divided at a declination of $-5^\circ$) can be seen in Table~\ref{tab:srclist}, where the best-fit number of astrophysical neutrino events $\hat{n}_s$ is constrained to be $\geq0$. For sources where $\hat{n}_s=0$, the 90\% C.L. median sensitivity was used in place of an upper limit. 

\begin{figure}[ht]
    \includegraphics[width=0.35\textwidth]{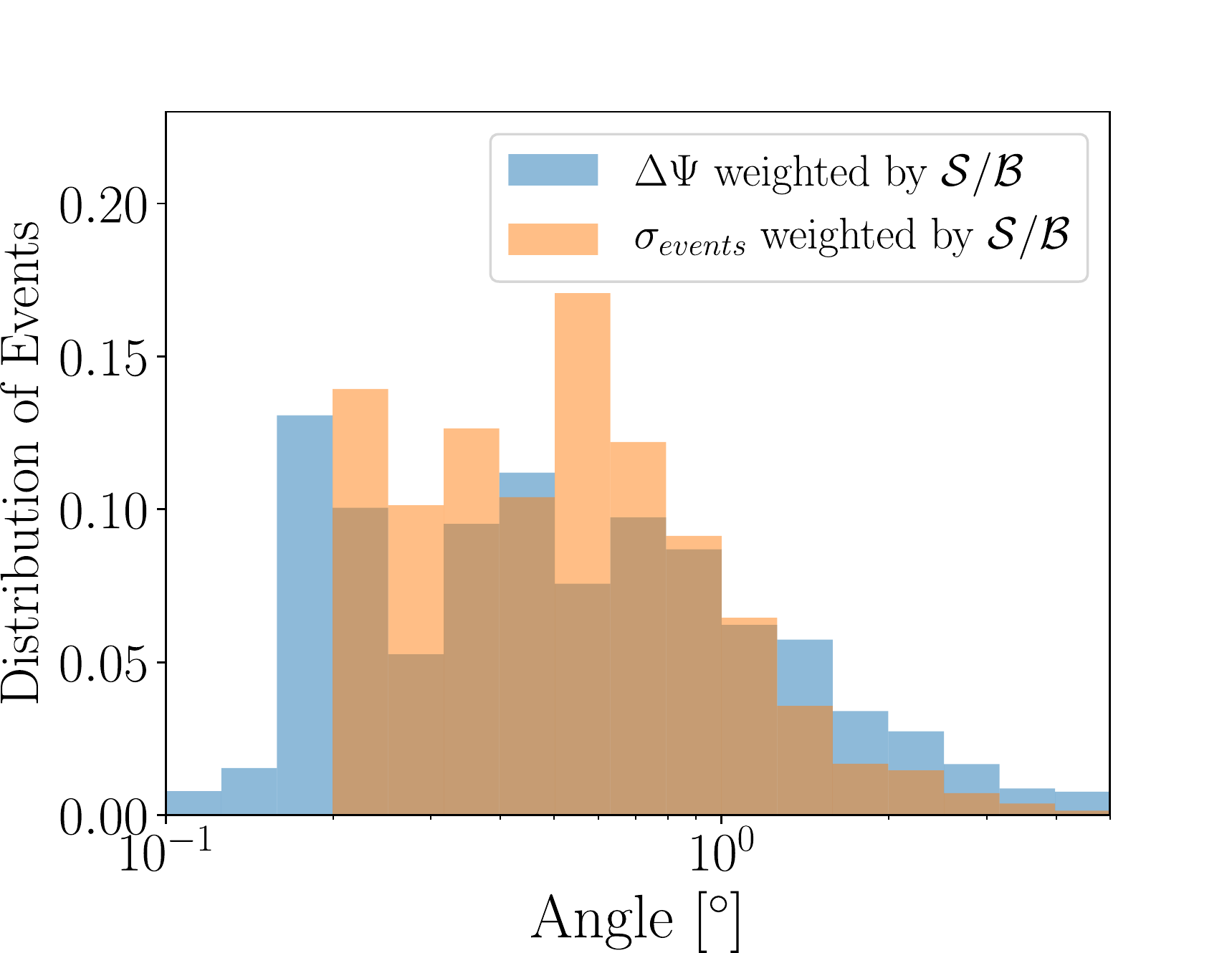}
     \caption{Real event distribution of the reconstructed angular uncertainty from the source (paraboloid~\cite{Neunhoffer:2004ha} $\sigma$ in orange) and the angular distances between NGC 1068 and each event($\Delta\Psi$ in blue), both weighted by their signal over background likelihood for a given point-like source hypothesis in the direction of NGC 1068 and the best fit spectral shape of $E^{-3.2}$. 
     }
    \label{fig:angErr}
\end{figure}

The most significant excess from the Northern Catalog was found in the direction of NGC 1068. Figure \ref{fig:angErr} shows the distribution of observed events as a function of their distance from the 3FGL coordinates of NGC 1068 (blue) or their estimated angular error (orange). Both distributions are weighted by their signal over background likelihood for a given point-like source hypothesis in the direction of NGC 1068 and the best fit spectral shape of $E^{-3.2}$. A minimum angular uncertainty of $0.2^\circ$ is applied because the angular uncertainty $\sigma$ estimated for each event individually does not include systematic uncertainties. It was verified that setting a minimum value up to $0.9^\circ$ does not significantly affect the result in the direction of NGC 1068 as most events contributing to the excess are reconstructed within $\sim1^\circ$ of the \textit{Fermi}-LAT NGC 1068 coordinates. 

\begin{figure}
    \centering
    \includegraphics[width=0.5\textwidth]{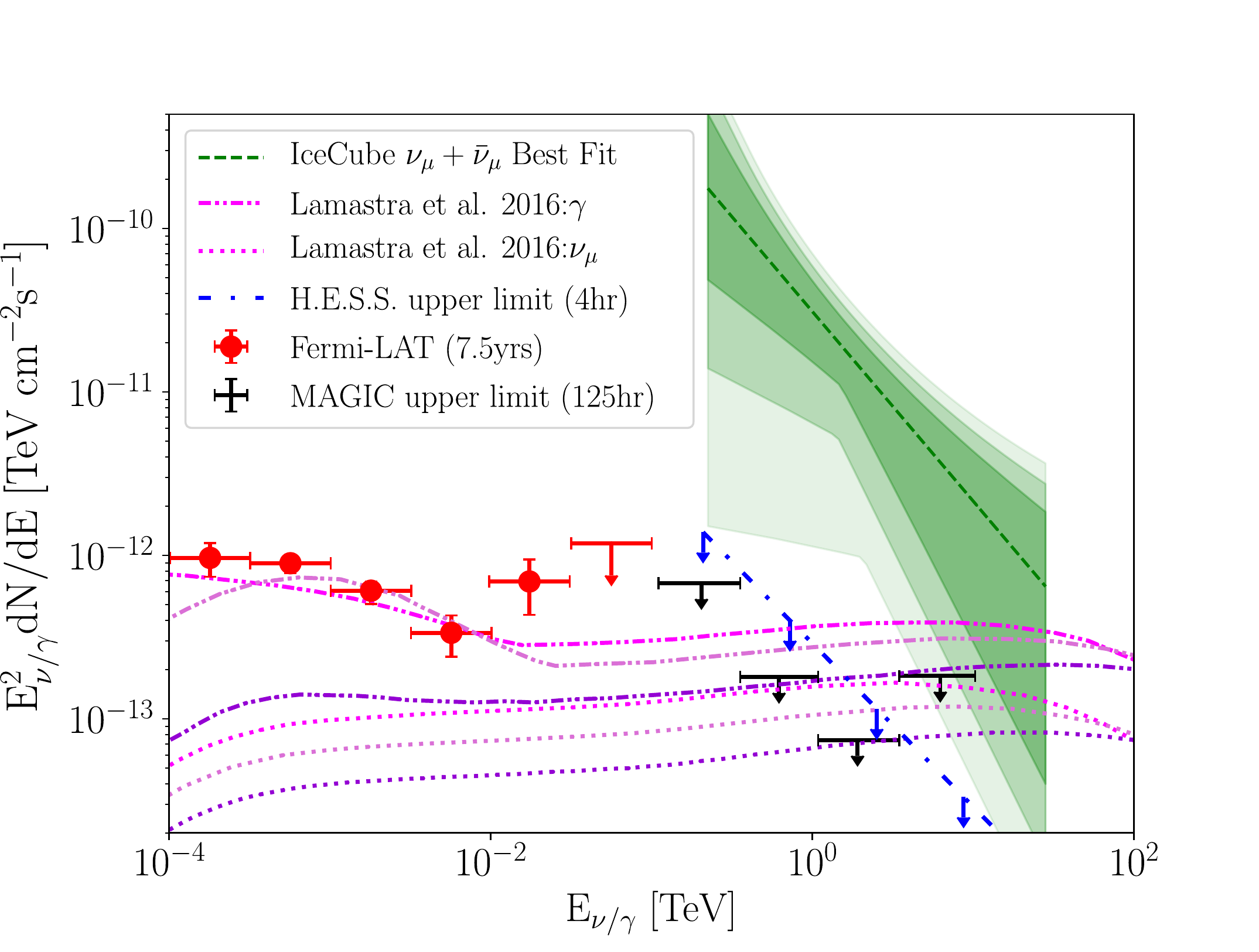}
    \caption{The best-fit time-integrated astrophysical power-law neutrino flux obtained using the 10 year IceCube event selection in the direction of NGC 1068. The shaded regions represent the 1, 2 \& 3$\sigma$ error regions on the spectrum as seen in Fig.~\ref{fig:gammaScan}. This fit is compared to the $\gamma$ and corresponding $\nu$ AGN outflow models and the \textit{Fermi} Pass8 (P8) results found in \citet{Lamastra:2016axo} (which do not include modelled absorption effects \cite{Lamastra:2017iyo}). AGN-driven outflow parameters are set at $R_{out}$=100\,pc, $v_{out}$=200\,km/s, $p=2$, and $L_{kin}$=1.5$\times10^{42}$\,erg/s; violet: $L_{AGN}$=4.2$\times10^{44}$\,erg/s, $n_H$=$10^{4}$\,cm$^{-3}$, $F_{cal}=1$, $\eta_p=0.2$, $\eta_e=0.02$, $B_{ISM}=30\,\mu$G; magenta: $L_{AGN}$=2.1$\times10^{45}$\,erg/s, $n_H$=120\,cm$^{-3}$, $F_{cal}=0.5$, $\eta_p=0.5$, $\eta_e=0.4$, $B_{ISM}=250\,\mu$G; pale pink: $L_{AGN}$=4.2$\times10^{44}$\,erg/s, $n_H$=$10^4$\,cm$^{-3}$, $F_{cal}=1$, $\eta_p=0.3$, $\eta_e=0.1$, $B_{ISM}=600\,\mu$G. The upper-limits in $\gamma$-ray observations are taken from from H.E.S.S. (blue)~\citet{Aharonian:2005ar} and from MAGIC (black)~\citet{Acciari:2019raw}. }
    \label{fig:ngc1068_mwl}
\end{figure}

Finally, to provide more context for such a result, we show the reconstructed muon neutrino spectrum with its large uncertainty compared to gamma-ray data from 7.5 yr of Fermi-LAT observations and an upper limit obtained from 125 hrs of MAGIC observations and about 4 hrs of H.E.S.S. observations \cite{Acciari:2019raw,Aharonian:2005ar,2012ApJ...755..164A} in Fig.~\ref{fig:ngc1068_mwl}. 

\begin{figure}
    \centering
    \includegraphics[width=0.35\textwidth]{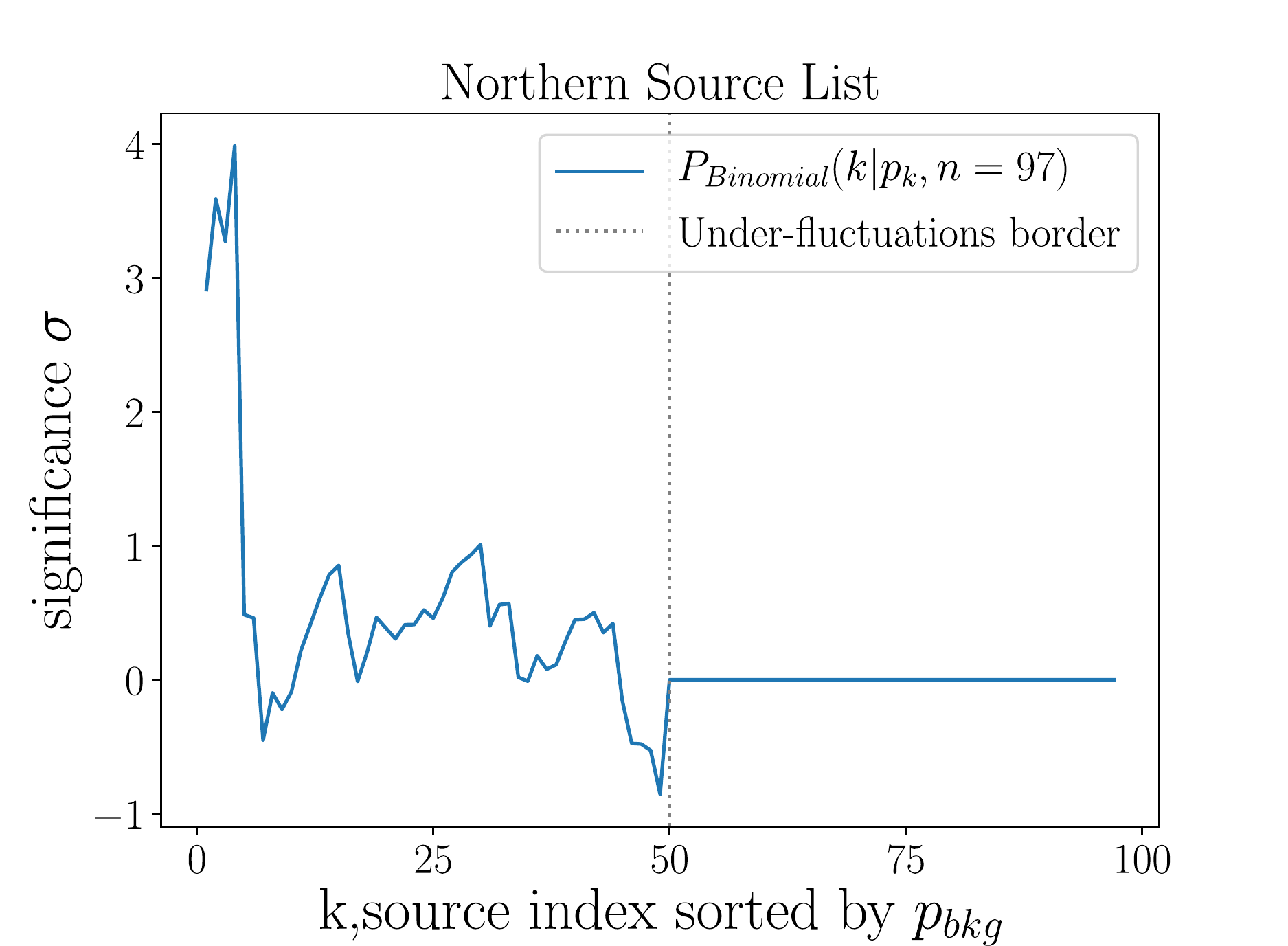}
    \includegraphics[width=0.35\textwidth]{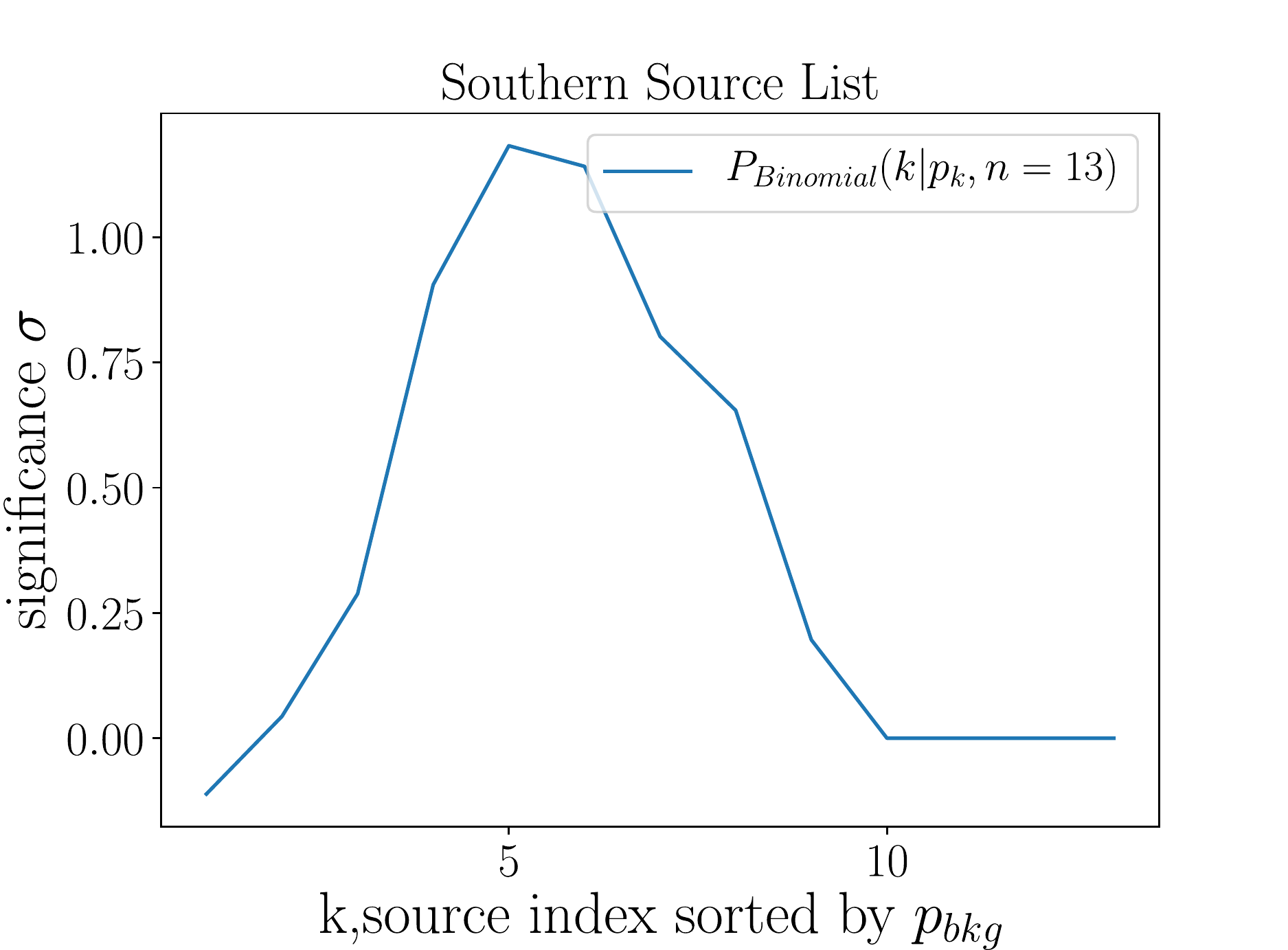}
    \caption{\textit{Left:} Significance of the pre-trial probability of obtaining $k$ excesses with the significance of the $k^{th}$ source or higher from the Northern catalog given background only. \textit{Right:} Equivalent plot for the Southern catalog.}
    \label{fig:pop}
\end{figure}
\renewcommand{\arraystretch}{1.0}
\begin{table}
 \caption{Galactic sources examined in the stacked searches in three catalogs: Supernova Remnants (SNR), Pulsar Wind Nebula (PWN), and Unidentified Objects (UNID). For each source: equatorial coordinates (J2000), and the relative source weight used for the analysis are given.  
 \label{tab:gallist}}
\begin{tabular}[c]{| c  c  c  c  c  |}
 \hline
 \multicolumn{5}{ c }{Stacking Catalogs}\\
 \hline
 Catalog & Name & $\alpha\,[\mathrm{deg}]$ & $\delta\,[\mathrm{deg}]$ & Weight\\
 \hline
 SNR & HESS J1614-518 & 243.56 & -51.82 & 2.80$\times10^{-1}$\\
& HESS J1457-593 & 223.70 & -59.07 & 1.47$\times10^{-1}$\\
& HESS J1731-347 & 262.98 & -34.71 & 1.40$\times10^{-1}$\\
& HESS J1912+101 & 288.33 & 10.19 & 7.13$\times10^{-2}$\\
& SNR G323.7-01.0 & 233.63 & -57.20 & 6.91$\times10^{-2}$\\
& Gamma Cygni & 305.56 & 40.26 & 6.35$\times10^{-2}$\\
& CTB 37A & 258.64 & -38.55 & 5.01$\times10^{-2}$\\
& RX J1713.7-3946 & 258.36 & -39.77 & 3.94$\times10^{-2}$\\
& HESS J1745-303 & 266.30 & -30.20 & 2.77$\times10^{-2}$\\
& Cassiopeia A & 350.85 & 58.81 & 1.89$\times10^{-2}$\\
& HESS J1800-240B & 270.11 & -24.04 & 1.82$\times10^{-2}$\\
& W 51C & 290.82 & 14.15 & 1.65$\times10^{-2}$\\
& HESS J1800-240A & 270.49 & -23.96 & 1.48$\times10^{-2}$\\
& SN 1006 & 225.59 & -42.10 & 1.20$\times10^{-2}$\\
& W28 & 270.34 & -23.29 & 9.06$\times10^{-3}$\\
& CTB 37B & 258.43 & -38.17 & 8.19$\times10^{-3}$\\
& Vela Junior & 133.00 & -46.33 & 4.88$\times10^{-3}$\\
& LMC N132D & 81.26 & -69.64 & 4.83$\times10^{-3}$\\
& IC 443 & 94.51 & 22.66 & 2.51$\times10^{-3}$\\
& SNR G349.7+0.2 & 259.50 & -37.43 & 1.50$\times10^{-3}$\\
& Tycho SNR & 6.34 & 64.14 & 8.83$\times10^{-4}$\\
& W 49B & 287.75 & 9.10 & 5.04$\times10^{-4}$\\
& RCW 86 & 220.12 & -62.65 & 2.54$\times10^{-6}$\\

\hline
PWN & HESS J1708-443 & 257.00 & -44.30 & 1.63$\times10^{-1}$\\
& HESS J1632-478 & 248.01 & -47.87 & 1.19$\times10^{-1}$\\
& Vela X & 128.29 & -45.19 & 1.06$\times10^{-1}$\\
& HESS J1813-178 & 273.36 & -17.85 & 6.91$\times10^{-2}$\\
& MSH 15-52 & 228.53 & -59.16 & 6.58$\times10^{-2}$\\
& HESS J1420-607 & 214.69 & -60.98 & 6.27$\times10^{-2}$\\
& HESS J1837-069 & 279.43 & -6.93 & 5.78$\times10^{-2}$\\
& HESS J1616-508 & 244.06 & -50.91 & 5.41$\times10^{-2}$\\
& HESS J1026-582 & 157.17 & -58.29 & 5.05$\times10^{-2}$\\
& HESS J1356-645 & 209.00 & -64.50 & 4.25$\times10^{-2}$\\
& PSR B0656+14 & 104.95 & 14.24 & 4.04$\times10^{-2}$\\
& HESS J1418-609 & 214.52 & -60.98 & 3.81$\times10^{-2}$\\
& HESS J1849-000 & 282.26 & -0.02 & 2.51$\times10^{-2}$\\
& Geminga & 98.48 & 17.77 & 2.26$\times10^{-2}$\\
& HESS J1825-137 & 276.55 & -13.58 & 1.90$\times10^{-2}$\\
& CTA 1 & 1.65 & 72.78 & 1.61$\times10^{-2}$\\
& SNR G327.1-1.1 & 238.63 & -55.06 & 8.37$\times10^{-3}$\\
& SNR G0.9+0.1 & 266.83 & -28.15 & 5.47$\times10^{-3}$\\
& SNR G054.1+00.3 & 292.63 & 18.87 & 5.11$\times10^{-3}$\\
& Crab nebula & 83.63 & 22.01 & 4.57$\times10^{-3}$\\
& HESS J1846-029 & 281.50 & -2.90 & 4.18$\times10^{-3}$\\
& SNR G15.4+0.1 & 274.50 & -15.45 & 3.99$\times10^{-3}$\\
& HESS J1119-614 & 169.81 & -61.46 & 3.49$\times10^{-3}$\\
& VER J2016+371 & 304.01 & 37.21 & 3.14$\times10^{-3}$\\
& HESS J1458-608 & 224.87 & -60.88 & 2.46$\times10^{-3}$\\
& HESS J1833-105 & 278.25 & -10.50 & 2.24$\times10^{-3}$\\
& N 157B & 84.44 & -69.17 & 1.52$\times10^{-3}$\\
& 3C 58 & 31.40 & 64.83 & 1.30$\times10^{-3}$\\
& HESS J1303-631 & 195.75 & -63.20 & 1.22$\times10^{-3}$\\
& DA 495 & 298.06 & 29.39 & 6.29$\times10^{-4}$\\
& HESS J1018-589 B & 154.09 & -58.95 & 3.22$\times10^{-4}$\\
& HESS J1718-385 & 259.53 & -38.55 & 2.56$\times10^{-4}$\\
& HESS J1640-465 & 250.12 & -46.55 & 1.56$\times10^{-5}$\\
\hline
\end{tabular}
\end{table}

\begin{table}
\begin{tabular}[c]{| c  c  c  c  c |}
 \hline
 \multicolumn{5}{ c }{Stacking Catalogs}\\
 \hline
 Catalog & Name & $\alpha\,[\mathrm{deg}]$ & $\delta\,[\mathrm{deg}]$ & Weight\\
 \hline
UNID & HESS J1702-420 & 255.68 & -42.02 & 1.80$\times10^{-1}$\\
& MGRO J2019+37 & 304.01 & 37.20 & 1.17$\times10^{-1}$\\
& Westerlund 1 & 251.50 & -45.80 & 1.04$\times10^{-1}$\\
& HESS J1626-490 & 246.52 & -49.09 & 5.91$\times10^{-2}$\\
& HESS J1841-055 & 280.23 & -5.55 & 5.60$\times10^{-2}$\\
& HESS J1809-193 & 272.63 & -19.30 & 5.07$\times10^{-2}$\\
& HESS J1843-033 & 280.75 & -3.30 & 4.80$\times10^{-2}$\\
& MGRO J1908+06 & 287.17 & 6.18 & 4.67$\times10^{-2}$\\
& HESS J1857+026 & 284.30 & 2.67 & 2.91$\times10^{-2}$\\
& HESS J1813-126 & 273.35 & -12.77 & 2.90$\times10^{-2}$\\
& 2HWC J1814-173 & 273.52 & -17.31 & 2.61$\times10^{-2}$\\
& HESS J1831-098 & 277.85 & -9.90 & 1.90$\times10^{-2}$\\
& HESS J1852-000 & 283.00 & 0.00 & 1.77$\times10^{-2}$\\
& HESS J1427-608 & 216.97 & -60.85 & 1.71$\times10^{-2}$\\
& TeV J2032+4130 & 308.02 & 41.57 & 1.64$\times10^{-2}$\\
& Galactic Centre ridge & 266.42 & -29.01 & 1.24$\times10^{-2}$\\
& HESS J1708-410 & 257.10 & -41.09 & 1.17$\times10^{-2}$\\
& VER J2227+608 & 336.88 & 60.83 & 1.05$\times10^{-2}$\\
& HESS J1634-472 & 248.50 & -47.20 & 1.00$\times10^{-2}$\\
& 2HWC J1949+244 & 297.42 & 24.46 & 9.92$\times10^{-3}$\\
& HESS J1834-087 & 278.72 & -8.74 & 9.65$\times10^{-3}$\\
& HESS J1507-622 & 226.88 & -62.42 & 9.57$\times10^{-3}$\\
& 2HWC J1819-150 & 274.83 & -15.06 & 9.36$\times10^{-3}$\\
& 2HWC J0819+157\footnotemark[1] & 124.98 & 15.79 & 8.48$\times10^{-3}$\\
& HESS J1641-463 & 250.26 & -46.30 & 7.72$\times10^{-3}$\\
& HESS J1858+020 & 284.58 & 2.09 & 7.56$\times10^{-3}$\\
& HESS J1503-582 & 225.75 & -58.20 & 7.31$\times10^{-3}$\\
& 2HWC J1040+308\footnotemark[1] & 160.22 & 30.87 & 7.14$\times10^{-3}$\\
& Westerlund 2 & 155.75 & -57.50 & 6.80$\times10^{-3}$\\
& HESS J1804-216 & 271.12 & -21.73 & 6.60$\times10^{-3}$\\
& 2HWC J1309-054 & 197.31 & -5.49 & 4.19$\times10^{-3}$\\
& HESS J1828-099 & 277.25 & -9.99 & 4.16$\times10^{-3}$\\
& 2HWC J1928+177 & 292.15 & 17.78 & 3.32$\times10^{-3}$\\
& HESS J1848-018 & 282.12 & -1.79 & 3.03$\times10^{-3}$\\
& HESS J1729-345 & 262.25 & -34.50 & 2.91$\times10^{-3}$\\
& 2HWC J1955+285 & 298.83 & 28.59 & 2.78$\times10^{-3}$\\
& 2HWC J1852+013 & 283.01 & 1.38 & 2.76$\times10^{-3}$\\
& 2HWC J2024+417 & 306.04 & 41.76 & 2.71$\times10^{-3}$\\
& 2HWC J2006+341\footnotemark[2] & 301.55 & 34.18 & 2.64$\times10^{-3}$\\
& HESS J1808-204 & 272.00 & -20.40 & 2.05$\times10^{-3}$\\
& 2HWC J1829+070 & 277.34 & 7.03 & 1.99$\times10^{-3}$\\
& Arc source & 266.58 & -28.97 & 1.99$\times10^{-3}$\\
& 2HWC J1921+131 & 290.30 & 13.13 & 1.69$\times10^{-3}$\\
& 2HWC J1953+294 & 298.26 & 29.48 & 1.65$\times10^{-3}$\\
& HESS J1832-085 & 278.13 & -8.51 & 1.55$\times10^{-3}$\\
& Terzan 5 & 267.02 & -24.78 & 1.54$\times10^{-3}$\\
& 2HWC J1914+117 & 288.68 & 11.72 & 1.51$\times10^{-3}$\\
& HESS J1741-302 & 265.25 & -30.20 & 1.49$\times10^{-3}$\\
& HESS J1844-030 & 281.17 & -3.10 & 1.33$\times10^{-3}$\\
& 2HWC J1938+238 & 294.74 & 23.81 & 9.80$\times10^{-4}$\\
& HESS J1832-093 & 278.19 & -9.37 & 9.22$\times10^{-4}$\\
& HESS J1826-130 & 276.50 & -13.09 & 9.21$\times10^{-4}$\\
& 2HWC J1902+048 & 285.51 & 4.86 & 6.17$\times10^{-4}$\\
& 2HWC J1907+084 & 286.79 & 8.50 & 5.08$\times10^{-4}$\\
& 30 Dor C & 83.96 & -69.21 & 3.07$\times10^{-4}$\\
& Galactic Centre & 266.42 & -29.01 & 1.83$\times10^{-4}$\\
& MAGIC J0223+403 & 35.67 & 43.04 & 9.46$\times10^{-5}$\\
& HESS J1746-308 & 266.57 & 30.84 & 7.88$\times10^{-5}$\\
\hline
\hline
 \end{tabular}
 \footnotetext[1]{Assumed extension of 2.0$^\circ$}
 \footnotetext[2]{Assumed extension of 0.9$^\circ$}
 \end{table}

\end{document}